# Quantifiably Tuneable Luminescence by Ultra-Thin Metal-Organic Nanosheets via Dual-Guest Energy Transfer

Dylan A. Sherman,[a] Mario Gutiérrez,[b] Ian Griffiths,[c] Samraj Mollick,[a] Nader Amin,[d] Abderrazzak Douhal,[b] and Jin-Chong Tan [a,*]

[a]Multifunctional Materials & Composites (MMC) Laboratory, Department of Engineering Science, University of Oxford, Parks Road, Oxford OX1, United Kingdom.

[b]Departamento de Química Física, Facultad de Ciencias Ambientales y Bioquímica, INAMOL, Universidad de Castilla-La Mancha, Toledo 45071, Spain.

[c]Department of Materials, University of Oxford, 16 Parks Road, Oxford OX1 3PH, United Kingdom.

[d]Department of Chemistry, University of Oxford, Mansfield Road, Oxford OX1 3TA, United Kingdom.

[*]Corresponding author: jin-chong.tan@eng.ox.ac.uk

## Abstract

Luminescent metal-organic frameworks (LMOFs) are promising materials for organic light-emitting diode (OLED) alternatives to silicate-based LEDs due to their tuneable structure and programmability. Yet, the 3D nature of LMOFs creates challenges for stability, optical transparency, and device integration. Metal-organic nanosheets (MONs) potentially overcome these limitations by combining the benefits of MOFs with an atomically thin morphology of large planar dimensions. Here, we report the bottom-up synthesis of atomically thin ZIF-7-III MONs *via* facile low-energy salt-templating. Employing guest@MOF design, the fluorophores Rhodamine B and Fluorescein were intercalated into ZIF-7 nanosheets (Z7-NS) to form light emissive systems exhibiting intense and highly photostable fluorescence. Aggregation and Förster resonance energy transfer, enabled by the MON framework, were revealed as the mechanisms behind fluorescence. By varying guest concentration, these mechanisms provided predictable quantified control over emission chromaticity of a dual-guest Z7-NS material and the definition of an 'emission chromaticity fingerprint' - a unique subset of the visible spectrum which a material can emit by fluorescence.

## Introduction

Metal-Organic Framework (MOF) nanosheets, recently termed Metal-Organic Nanosheets (MONs), are an increasingly sought-after form of MOFs that unite the properties of 2D materials with 3D MOF characteristics.[1,2] Nanosheets are high surface-to-volume atom ratio materials with single or few-atom thickness (1-10 nm typically) and comparatively large micron-scale lateral planar dimensions.[3] Archetypal 2D materials, such as MXenes,[4] layered double hydroxides (LDH) and oxides,[5] and graphitic carbon nitrides,[6] have been reduced to nanosheets in multiple studies.[7] These nanosheets have unprecedented physical, electronic, chemical, and optical properties unattainable in their 3D layered bulk counterparts.[7] Challengingly, restricting material size in one or more dimensions may limit long-range effects, functionality, and the structural diversity afforded by stacking 2D sheets into bulk 3D materials.[7] MONs provide a competitive solution to these nanosheet materials by adding key MOF characteristics derived from the metal-organic composition. These include structural diversity and tuneability, programmable functionality, mechanical anisotropy,[8,9] and highly ordered pore arrays with abundant accessible active sites.[1] To date, MONs have shown great promise in both gas separation and water purification applications,[10,11] energy storage,[12] light harvesting and emission,[1] electronic devices,[13] catalysis,[14,15] and sensing.[16]

The study of MONs is still in its infancy, with synthesis and stability often the greatest challenges.[3,17] Most studies rely on top-down disassembly of 2D layered MOFs,[18,19] which is energy and time consuming, and typically result in MONs of non-uniform nanosheet dimensions with rough fragmented surfaces that commonly reaggregate.[18] Bottom-up synthesis provides greater morphological control and stable isolated sheets, but with few reports in high yield.[14] Examples of bottom-up techniques tailored to specific MOFs, such as ZIF-8 and ZIF-67, include modulation,[20,21] diffusion,[22] surfactant assistance,[23,24] salt-confinement,[25] and interfacial synthesis.[26] ZIF nanosheets are particularly intriguing, given that 3D ZIFs exhibit high stability, hydrophobicity, and large pore size.[27] ZIF-7 [$Zn(bIm)_2$, bIm = benzimidazolate] is isolatable in a dense thermodynamically stable 2D-layered phase ZIF-7-III ($Zn_2(bIm)_4$),[28] making it viable for the formation of nanosheets (Fig. 1).[29] ZIF-7 nanosheets have been used as a matrix for the ionisation of small molecules,[30] molecular sieving,[31] and membranes.[31] To our knowledge, though, there are no reports of ZIF-7-III nanosheet bottom-up synthesis, only top-down delamination.[26,30,32,33]

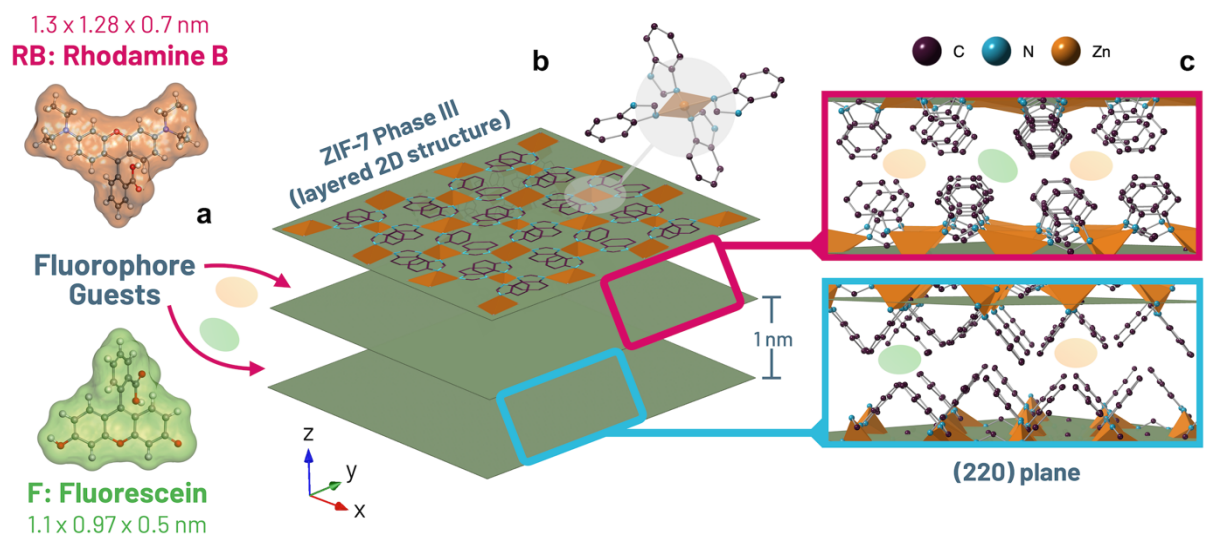

**Figure 1.** Schematic of material design with structures from single crystal ZIF-7-III data.[33] (a) Chemical structures of Rhodamine B and Fluorescein. (b) Single secondary building unit with ZnN₄ tetrahedra highlighted on a single sheet (monolayer) of $Zn_2(bIm)_4$ in the *xy*-plane of a segment of ZIF-7-III layered structure with green planes indicating the (002) *d*-spacing. (c) Cross section of the (220) plane, illustrating the perpendicular arrangement of layers and interlayer spacing. For all representations: carbon (purple), nitrogen (blue), Zn(II) (orange), oxygen (red), hydrogen (omitted for clarity).

One of the lesser explored areas of MONs is lighting. A fluorescent nano-emitter should have negligible dye leaching, structural stability under ambient conditions and repeated use, and high photoluminescence quantum yield (PLQY). The majority of luminescent MOF (LMOF) literature targets 3D bulk materials,[35] but these are often unstable in solution due to guest leaching. The host framework can also impede emission *via* self-quenching, either through energy interconversion and absorption or physical shielding.[36] The increased specific surface area and atomic thickness of MONs, contrastingly, improves optical transparency while allowing for strong in-plane covalent bonding for mechanical strength and flexibility.[37,38,9] This has resulted in MONs being typically highly dispersive and stable in water or solvents.[3] The confinement of electrons in an ultra-thin region also facilitates the control and directionality of excitons emission, which theoretically allows for the optimisation of guests luminescence.[39] Exfoliated materials such as AUBM-6-NS, for example, have reported up to threefold greater emission intensity than the bulk AUBM-6.[37]

At present, the overwhelming majority of luminescent MONs (LMONs) utilise inherent luminescence *via* the design of emissive linkers, metal ions, or *via* ligand-metal charge transfer.[40] A novel route to LMONs is employing the guest@MOF approach and, to the best of our knowledge, the only work to date considering this approach is the system lanthanide hydrate@MOF nanosheets (HSB-W5-NS) *via* metal ion guest@MOF.[41] Dyes, alternatively, are ideal guests used in guest@MOF systems.[42] Rhodamine B (RB) and Fluorescein (F) (Fig. 1a) are two of the most employed guests due to their high PLQY and stability.[43] While F and RB do not fluoresce in the solid-state form due to aggregation-caused quenching (ACQ),[44,45] isolating monomers *via* encapsulation in pores of porous superstructure hosts, such as an array of 3D MOFs,[46-49] can induce fluorescence in solid state.[50,51] In a 2D MON system, we theorise this encapsulation is analogous to intercalation, which has been achieved for RB and F in LDH systems and clays,[52,53] yet the resulting systems are frequently reported as being unstable or lack tuneability.[54,55] The intercalation of organic fluorophores is yet to be explored in MONs, despite the apparent advantages of a more flexible and customisable parent matrix to control loading and tune interlayer spacing.

Herein, we report the first *in situ* bottom-up synthesis of ultrathin $Zn_2(bIm)_4$ MONs (Z7-NS) *via* a low-energy and scalable salt-template synthesis. We show that the fluorescent dyes RB and F can be incorporated into individual Z7-NS using this synthesis to yield highly stable fluorescent guest@MON systems with tuneable emission of visible light. Utilising nanoscale analysis techniques we reveal the underpinning mechanisms of guest intercalation. When the dyes are intercalated simultaneously, a dual-guest yellow emitting nano-system results, whose emission chromaticity is precisely tuneable *via* numerical modelling.

## Results and Discussion

### Synthesis and Structure of Z7-NS Guest@MON Materials

Z7-NS were synthesised *in situ* using an environmentally benign adapted salt-templating technique during a one-pot reaction (see Methods and Supplementary Fig. 1).[25,56,57] By adding F or RB dissolved in methanol into the synthesis solution, facile *in situ* guest incorporation into the framework was achieved. Two single-guest systems (F@Z7-NS and RB@Z7-NS) at various guest loadings (guest@Z7-NS-$10^{-1}$, -$10^{-2}$ and -$10^{-3}$ M using synthesis guest quantities of 0.03, 0.003 and 0.0003 mmol respectively) were formed, along with a dual-guest system (entrapping F and RB simultaneously, DG@Z7-NS). Guest loading was quantified using $^1$H NMR (see Supplementary Fig. 2 and Table 1) with data confirming guest loading increases with an increase in guest concentration during synthesis. Synthesis was scalable, tested up to 10 times the initial quantities, producing solid material yields from 93 mg to 1.22 g (70-80%). Z7-NS remained as stable dispersions in $H_2O$ or MeOH, confirmed by Tyndall scattering (Supplementary Fig. 3), and are thermally stable up to 550 °C (Supplementary Fig. 4).

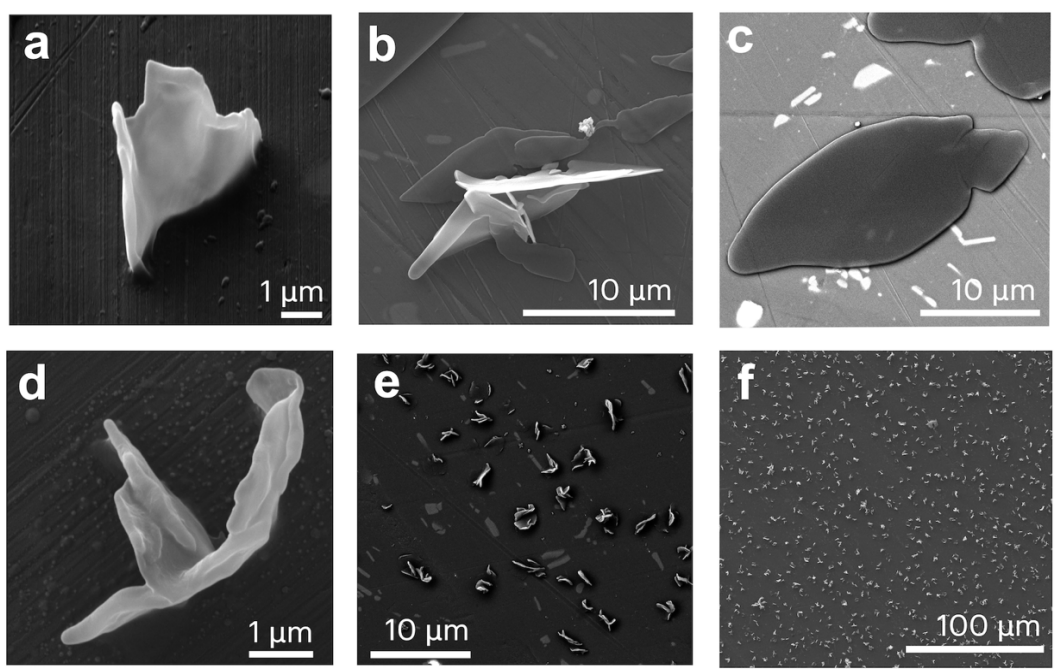

**Figure 2.** (a-f) FE-SEM secondary electron micrographs of Z7-NS collected at 10 keV and working distance of 9 mm. (a-c) primary nanosheet morphology showing homogenous dispersions and 3D thinness at 45° tilt (a); (d-f) secondary curled morphology shown from identical sample field area imaged at 3 different zoom levels.

FE-SEM imaging revealed Z7-NS and Guest@Z7-NS materials formed consistently in two morphologies, separable by gravitational sedimentation (Fig. 2). First, plate-like nanosheets with lateral dimensions of ~20×10 $\mu m^2$ (Fig. 2a-c) and a typical layer height, measured by AFM, of 2 nm for Z7-NS (Supplementary Fig. 5), 3 nm for RB@Z7-NS (Supplementary Fig. 6), 3.5 nm for F@Z7-NS (Supplementary Fig. 7), and 4 nm for DG@Z7-NS (Supplementary Fig. 8), with aggregates of at most 3 monolayers (~6-9 nm) (*c.f.* ZIF-7-III 2D layer stacks of 200 nm average height) (Supplementary Fig. 9). Second, curled nanosheets with dimensions of ~5×2 $\mu m^2$ (Fig. 2d-f). Similar curling has been observed by silica nanosheets from mechanical strain while mixing during synthesis. [58] SEM micrographs of Z7-NS imaged at a 45° tilt highlighted the thinness, curvature, and isolated nature of the nanosheets (Fig. 2d, Supplementary Fig. 10). All nanosheets were seen to be dispersed homogenously across the sample slide, on which they were drop casted for microscopic examinations (Fig. 2e-f). The morphology remains comparable after 6 months of immersion in water or MeOH, with only slightly increased curling at the nanosheet edges (Supplementary Fig. 11).

The nanosheet morphology and guest incorporation do not affect the long-range periodicity growth of the host framework, with ZIF-7-III like crystal structure confirmed by direct alignment of PXRD pattern reflections (Fig. 3a), IR and Raman spectroscopic bands (Supplementary Figs. 12-13). There is no indication of structural transformation or other phase impurities such as NaCl (Supplementary Fig. 14). Diffraction data further indicate the ultrathin 2D nature of the Z7-NS and the retention of crystallinity across the thin 2D plane. In PXRD patterns (Fig. 3a), peak broadening compared to ZIF-7-III and significant reduction in the relative peak intensity ratio of $I(002)/I(220)$ reflections from 14.814 in ZIF-7-III to 3.941 in ZIF-7-NS indicate reduced diffraction in the *z*-axis (out-of-plane orientation, see Fig. 1b) of the Z7-NS. HR-TEM (Fig. 3b-c, Supplementary Fig. 15) shows large domains of consistent *d*-spacing across a single nanosheet with an interlayer spacing of 0.56 nm, corresponding to the (220) planes, and 0.25 nm within each of the (220) layers, corresponding to the (260) planes. In the SAED pattern (Fig. 3d), three distinctive planes are identifiable upon indexing: the (220) (lowest angle (*hkl*) reflection with no symmetry about the *z*-axis), (110) and (440). All three planes have no *z*-component of symmetry. Thus, we posit the nanosheet has minimal atomic density in the *z*-axis, while the *xy*-plane is significantly atomically dense to produce intense diffraction.

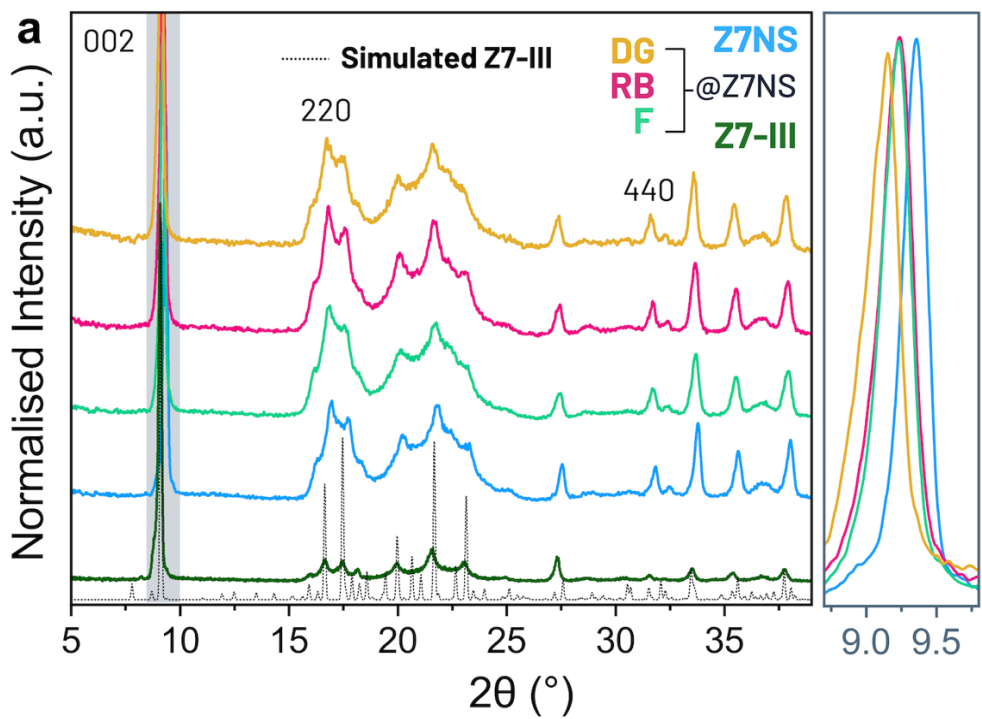

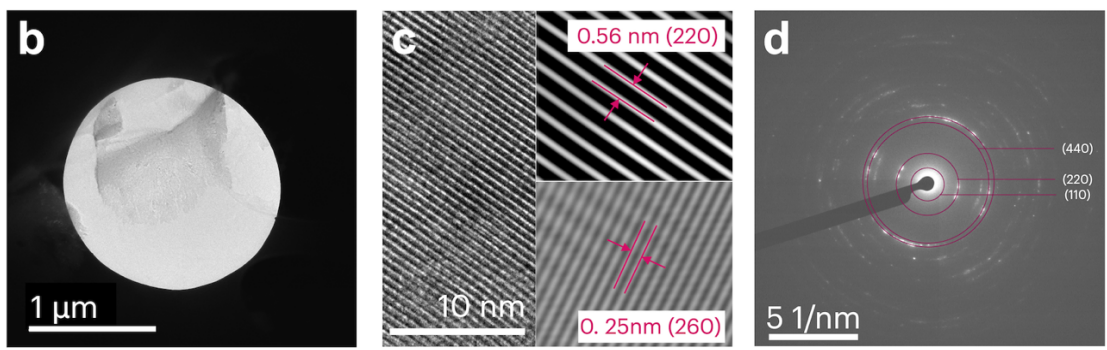

**Figure 3.** (a) PXRD patterns of Z7-NS with various dye@Z7-NS. (b) HR-TEM image of a single Z7-NS particle studied in (c)-(d). (c) Fourier-transform analysed HR-TEM of Z7-NS revealing inter-plane (220) spacing and intra-plane spacing in the (260) plane. (d) Selected area electron diffraction (SAED) image of Z7-NS showing indexed planes.

With the ZIF-7 host framework clearly defined, we next considered the location of the guest molecules within the structure. It was not possible to convert the guest@ZIF-7 systems to ZIF-7-III, nor was post-synthesis guest diffusion successful in incorporating guests into ZIF-7-III stacked materials or Z7-NS. Instead, guests were only incorporable *via in situ* syntheses, suggesting the guests are in some way being nanoconfined or intercalated within the layered nanosheet system during its formation, rather than merely attached to the external surfaces. Sterically, each guest comprises a xanthene ring with appended phenyl group, known to rotate to a planar conformation in confined spaces (Fig. 1a).[59] This, along with their molecular dimensions (10.191 x 9.669 x 6.43 Å for F and 13.683 x 13.233 x 6.97 Å for RB),[48] make the guests feasible to incorporate into the Z7-NS host framework. A single layer of ZIF-7-III comprises of a (4,4) square planar grid formed by corner-sharing networks, quadruply linked, of $ZnN_4$ tetrahedra (Fig. 1b). Each monolayer, of ~0.5 nm thick, stacks orthogonally rotated on the z-axis via C-H/π interactions, producing 2D channels of ~1 nm in height through the material's (220) plane (Fig. 1c). These channels could accommodate F or RB guests, oriented either perpendicularly or at least inclined at an angle to the layers of Z7-NS. Alternatively, a gap of 0.5 nm between layers provides opportunity for parallel dye encapsulation, with guest phenyl rings able to rest within the pockets between the sterically bulky bIm ligands (these protrude from the Zn centres of each layer but have a degree of rotational flexibility to accommodate guests).

PXRD shows that the (002) d-spacing between nanosheet layers increased following guest intercalation, from 9.44 Å in Z7-NS to 9.57 Å in F@Z7-NS and RB@Z7-NS, and to 9.66 Å for DG@Z7-NS (Fig. 3a). As described earlier, layer height also increased in AFM surface profiles. Vibrational spectroscopy (Raman, MIR and FTIR) revealed no new bands, suggesting no new covalent bonds have formed between the guests and the framework via surface absorption or within the framework (Supplementary Fig. 12-13). Nor are changes observable for the Z7-NS bands, indicating guests are incorporated in a way that does not affect the surrounding host framework. Nano-FTIR (Fig. 4b-c, Supplementary Fig. 16) was used to probe the local nanoscale chemical composition of the guest@Z7-NS systems at single points and across 2D fields, with a spatial resolution of ~20 nm. Near-field vibrational data show the presence of several IR bands, in particular the strong band at 760 $cm^{-1}$ (corresponding to bIm aromatic ring in-plane and out-of-plane deformations),[59] that align with the Z7-NS nano-FTIR spectra and the corresponding materials' ATR-FTIR spectra. There are, however, no strong bands present that correlate to the spectra of F and RB (Fig. 4b-c, Supplementary Fig. 16). As previous studies in our group have demonstrated,[60] this finding indicates a lack of aggregated dye molecules on the MOF surface, or that the dyes are confined in a single pocket, but rather distributed throughout the nanosheets.

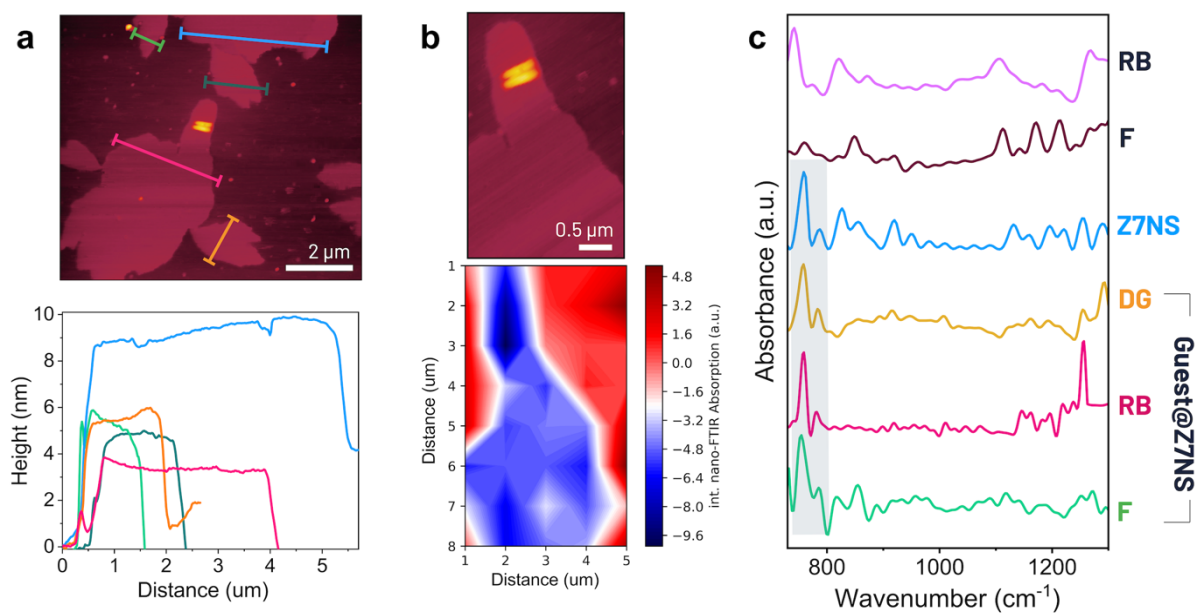

**Figure 4.** (a) AFM topography image (above) of DG@Z7-NS with height profiles (below) marked by the correspondingly coloured lines on the image. (b) Hyperspectral nano-FTIR imaging across a 2D region of DG@Z7-NS (see in AFM image above) between 725-1271 cm$^{-1}$. (c) Nano-FTIR spectra of various Z7-NS samples.

Notably, guest@Z7-NS samples showed negligible dye leaching when suspended in MeOH for 6 months, evincing protected encapsulation in the Z7-NS framework. Moreover, to ensure the presence of the dyes within the nanosheets, we have conducted fluorescence lifetime imaging microscopy (FLIM) experiments for the single guest@Z7-NS samples (Supplementary Fig. 17). The FLIM images show a good distribution of the fluorescent dyes over the whole nanosheets, with the emission spectra and decay lifetimes ($\tau$) comparable to that obtained for the bulk samples (*vide infra*). This local scale characterisation unequivocally proves the presence of the dyes in the nanosheet. Furthermore, from the nanoFTIR experiments it was evidenced the lack of the dyes on the external sample surface, and therefore, it substantiates the notion that the (guest) fluorophores are intercalated within the (host) nanosheet layers, entrapped within the framework or a combination of both.

## Photophysical Properties

The incorporation of dyes into the Z7-NS host enabled turn-on fluorescence of the guest in the solid state, with no indication of emission from the Z7-NS framework (Supplementary Fig. 18) or energy transfer between the LMON and guest (Fig. 5a-b). The emission spectral maps of F@Z7-NS and RB@Z7-NS (Supplementary Fig. 19) systems exhibit a single emission band characteristic of F and RB. A red shift of the emission is seen with increasing guest loading, resulting in tuneable chromaticity as a function of guest loading (Supplementary Fig. 20). This is an indication of guest aggregation as shown in other LG@MOFs systems.[43,47]

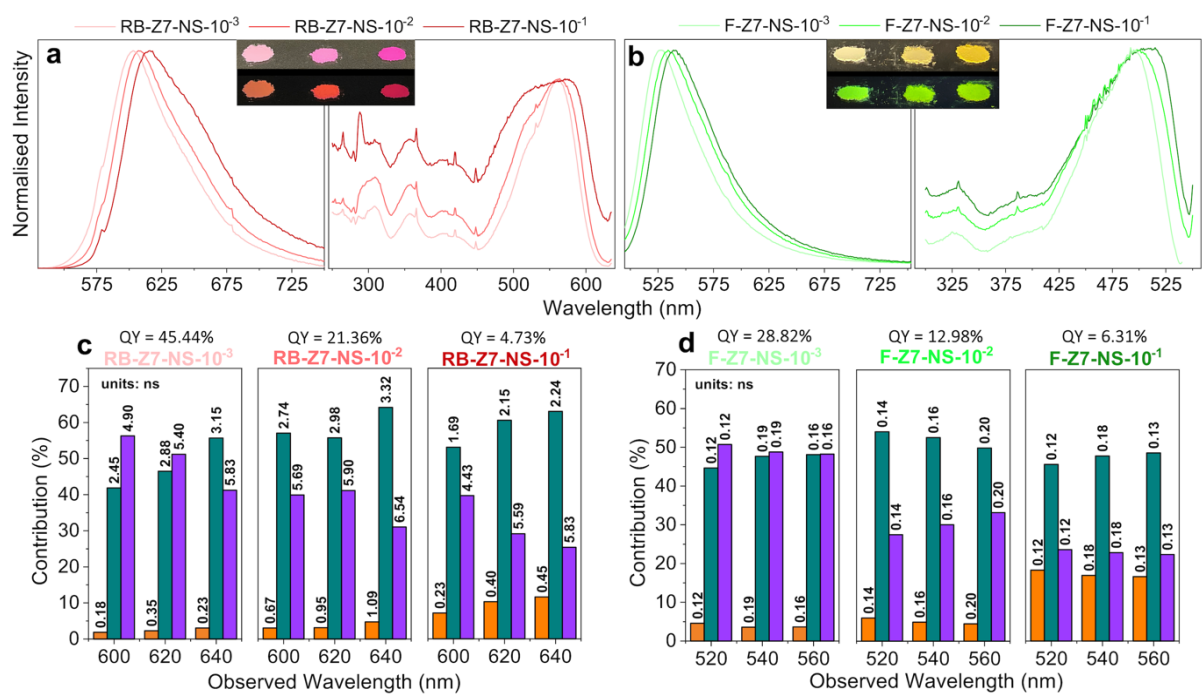

**Figure 5.** (a-b) Emission (left) spectra of guest@Z7 materials excited at 550 nm (RB) and 485 nm (F), and excitation spectra (right) of guest@Z7-NS materials with emission of 650 nm (RB) and 550 nm (F). Inset: powders of guest@Z7-NS-10⁻³, guest@Z7-NS-10⁻² and guest@Z7-NS-10⁻¹ (left-right) under ambient lighting (top) and UV light (bottom). (c-d) TCSPC fluorescence lifetime data and photoluminescence quantum yield (PLQY) of guest@Z7-NS materials across 3 observed wavelengths showing the percentage contribution $c$ on vertical axis and $\tau$ values above columns. Orange ($\tau_1$) = H-aggregates, green ($\tau_2$) = J-aggregates, and purple ($\tau_3$) = monomer species.

Excitation spectra and time-resolved photobehaviour obtained through picosecond time-correlated single photon counting (TCSPC) corroborate that guest molecule aggregation is a driver of these luminescent properties (Fig. 5c-d, Supplementary Tables 2-3). The fluorescence decays (observed at three wavelengths, upon excitation at the maximum of the absorption intensity) of F@Z7-NS and RB@Z7-NS were fitted using a three-component exponential function giving three time constants ($\tau_1$= 0.16 ns, $\tau_2$= 1.23 ns, and $\tau_3$= 3.76 ns for F@Z7-NS at 540 nm; and $\tau_1$= 0.95 ns, $\tau_2$= 2.98 ns, and $\tau_3$= 5.90 ns for RB@Z7-NS at 620 nm) that we assign to H-aggregates ($\tau_1$) (face-to-face), J-aggregates ($\tau_2$) (head-to-tail), and monomers ($\tau_3$) of F and RB, respectively. Our assignment is based on previous findings for guest@MOF systems and agrees with excitons theory.[46,47,61] Upon increasing concentration of the guest molecules, there is a reduction in contribution of the $\tau_3$-component (monomers), while the $\tau_2$-component (J-aggregates) exhibits minor increase proportional to guest loading. This observation suggests an increment in J-aggregates population, which is causing a red-shift in the absorption and emission spectra as explained above. On the other hand, $\tau_1$-component (H-aggregates) increases its contribution significantly with the guest loading, reflecting a major formation of H-aggregates. This is further supported by the growth of a shoulder band in the excitation spectrum at higher energies (470 nm for F and 515 nm for RB), and the reduction of PLQY, given emission from H-aggregates is theoretically forbidden.

These data suggest a mix of monomers and J-aggregates are preferred in dilute guest@Z7-NS samples, likely packing horizontally between the sheets given our structural analysis. With increased guest concentration, more efficient packing between layers leads to aggregation, with H-aggregates being more densely packed. Moreover, our structural analysis revealed the possibility of angular guest stacking and slipped cofacial forms of the J- and H-aggregates are well reported.[62] However, a saturation point appears with RB@Z7-NS-$10^{-1}$ evincing a large increase of H-aggregation. The emission spectra of RB@Z7-NS also present a shoulder around 650 nm that increases in intensity with higher guest loading, most notably from the $10^{-2}$ to $10^{-1}$ samples.

Importantly, we theorise this aggregation occurs in chains, producing long-range order resulting in larger orbitals overlap, as seen in the reduced band gap caused by increasing guest loading (Supplementary Fig. 21) and reported for other guest@MOF materials.[47] This phenomenon is more pronounced in the RB samples where emission decays are dependent on the gated spectral region, indicative of extended aggregate interactions, likely due to the molecular size of RB. Interestingly, while both RB@Z7-NS and F@Z7-NS exhibit solvatochromism in a range of common organic solvents (Supplementary Fig. 22), solvatochromic shifts are more pronounced in F@Z7-NS (Supplementary Fig. 23-24). The tighter aggregate packing and molecular size of RB may sterically hinder solvent interactions with RB guests, while in F the looser packing provides interaction sites for sensing — showing selectivity for aprotic solvents (*via* red-shift of emission) and polar protic solvents (higher emission intensity).

The photostability of these materials was also tested, both exhibiting good performance (Supplementary Fig. 25). After 24 hours of intense exposure to the spectrofluorometer's 150 W xenon bulb irradiating at a wavelength corresponding to each materials absorption maximum, RB@Z7-NS lost 10.5% intensity and F@Z7-NS lost 11.6% but retained chromaticity. 90% of this decrease occurred in the first 10 hours, with the following 14 hours plateauing to a stable intensity. This loss is comparable to F encapsulated in ZIF-8 reported materials, while F@ZIF-8 materials that exhibited surface aggregation saw over 35% intensity loss after 15 hours.[46] This shows a great promise for long-term light-emitting device use of the materials.

## Dual-Guest Emission Mechanism

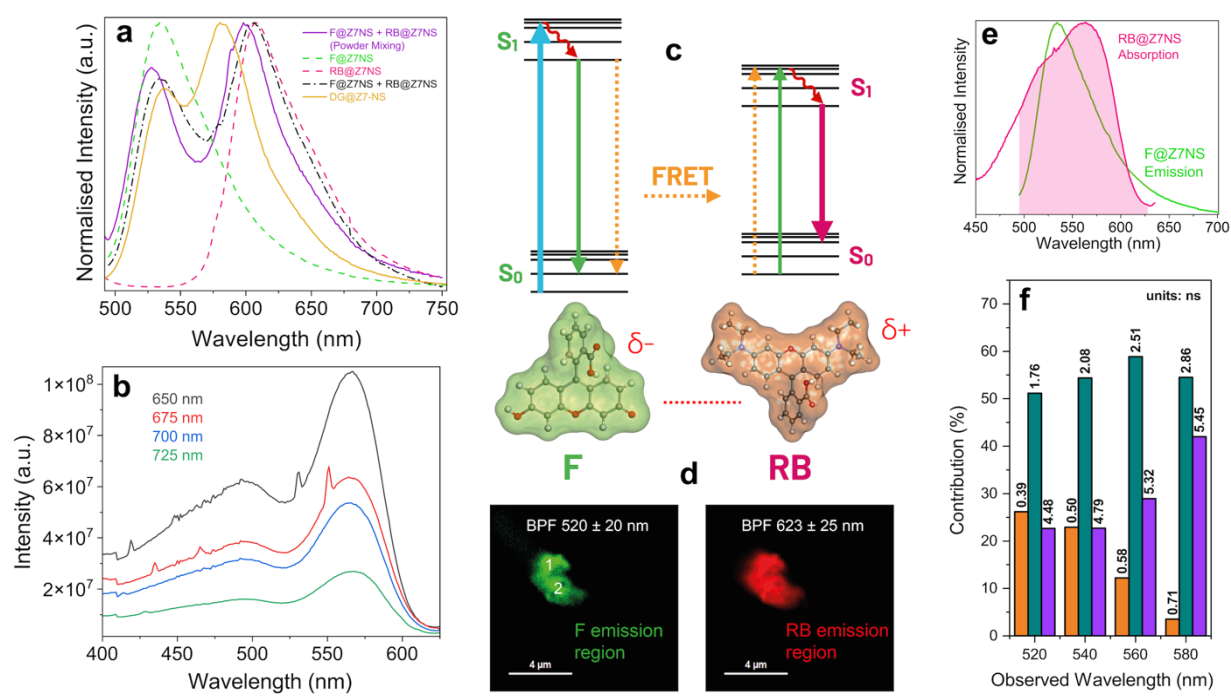

**Figure 6.** (a) Emission spectra of DG@ZIF-7-NS (DG = F+RB) compared to various spectra related to the single-guest F or RB@Z7-NS systems. Black curve indicates mathematical superposition of F@Z7-NS + RB@Z7-NS emission spectra. (b) Various excitation spectra of DG@Z7-NS recorded at indicated emission wavelengths. (c) Schematic of FRET between F and RB dipole-dipole bound dimers in DG@Z7-NS. (d) FLIM of DG@Z7-NS with two band pass filters (BPF) applied (left: 520 ± 20 nm and right 623 ± 25 nm). (e) Absorption spectra of RB@Z7-NS overlapped with emission spectra of F@Z7-NS. (f) Emission decay lifetimes. Orange ($\tau_1$) = FRET mechanism, green ($\tau_2$) = F, and purple ($\tau_3$) = RB.

By intercalating F and RB together in Z7-NS, at a total guest concentration below the determined saturation level for both guests, an optimised warm-yellow emitting material (DG@Z7-NS; DG = F+RB) was synthesised. Luminescence was achieved by exciting the material at the absorption intensity maximum of F (495 nm), producing an emission spectrum of two distinct but overlapping bands attributable to F (band at 525 nm) and RB (band at 580 nm, Fig. 6a). The excitation spectra of DG@Z7-NS present two bands with intensity maxima at 495 nm (attributed to the absorption of F) and 560 nm (due to the absorption of RB dye) (Fig. 6b and Supplementary Fig. 26). Remarkably, when the excitation spectrum is recorded at 725 nm (the emission of F dye is barely existent at this wavelength) the band at 495 nm is still observable, suggesting the occurrence of Förster resonance energy transfer (FRET, Fig. 6c) from the F to RB molecules within the Z7-NS sheets.[63] Theoretically, energy transfer is possible between F and RB guests due to the spectral overlap of the emission and absorption bands of F and RB, respectively (in the 520−540 nm region) (Fig. 6e).[64] FRET is rarely observed between guest dyes in MOF systems, however, due to the lack of proximity between dye molecules, which are typically entrapped within individual pores. The tight space within the MON layers, contrastingly, favours a short distance between guest dyes, triggering the FRET mechanism.[65] Resultingly, the energy absorbed by F upon excitation is partially interconverted to the RB system, from where it is then emitted simultaneously, to produce a yellow emission; the combination of the emission from F monomers not undergoing FRET and RB (Fig. 6).

Indeed, the fluorescence decays of bulk DG@Z7-NS were analysed by a sum of 3-components, which we assigned as follows: RB ($\tau_3$), F ($\tau_2$) and F-RB coupled FRET mechanism ($\tau_1$) (Fig. 6f, Supplementary Table 4). $\tau_3$ and $\tau_2$ are assigned as such for two reasons. This firstly corresponds with the single-guest system monomer component lifetimes (RB@Z7-NS > F@Z7-NS). Notably, though, the emission lifetime of F in the dual guest system ($\tau_2$) is largely quenched in comparison to F@Z7-NS, while $\tau_3$ is comparable to RB@Z7-NS, strong evidence of the proposed FRET mechanism. Secondly, as the observed wavelength varies from the maximum emission of F (~520 nm) to RB (~580 nm), the contribution of the $\tau_3$-component increases significantly, reflecting that this component is the lifetime of RB guest. $\tau_1$ is greater than the typical 0.1 ns observed for the H-aggregates of the single-guest Z7-NS systems.[47] Furthermore, the contribution of $\tau_1$ rapidly decreases as the observed wavelength increases, and therefore, it is sensible to attribute this contribution to the FRET mechanism, as it is demonstrated below for the DG@Z7-NS single particles.

FLIM experiments were also conducted to shed light on the photophysical mechanism of the DG@Z7-NS at a single crystal level. The images of several isolated MONs were collected using two band pass filters (BPF) at $520 \pm 20$ nm and $623 \pm 25$ nm to selectively record the emission of F and RB within the MONs, respectively (Fig. 6d and Supplementary Fig. 27). Remarkably, the fluorescence of both dyes is homogenous across the entire Z7-NS, illustrating the homogenous distribution (within the spatial resolution of the microscope, about 250 nm) of the two guests throughout the ultra-thin nanosheet system. The emission spectra collected at different individual points of these nanosheets closely match with the bulk observation, with two emission bands corresponding to F and RB fluorescence, confirming the phenomena identified in spectrofluorometer measurements, and which are originating from the Z7-NS samples rather than other impurities or crystallised fluorophore mixed phases. Moreover, the time-resolved photobehavior of the isolated thin films also aligns with that of the bulk samples, being the emission decays analyses as a sum of three exponentials ($\tau_1 = 480-520$ ps, $\tau_2 = 1.8-2.0$ ns, and $\tau_3 = 4.6-5.0$ ns when recording at $520 \pm 20$ nm, and $\tau_1 = 70-90$ ps, $\tau_2 = 1.9-2.2$ ns, and $\tau_3 = 4.4-5.1$ ns when gating at $623 \pm 25$ nm). Like our previous attribution, the longest $\tau_3$-component, whose contribution is higher at longer wavelengths ($623 \pm 25$ nm BPF), is assigned to the emission lifetime of the trapped RB dye. On the other hand, the $\tau_2$-component, contributing more at the bluest region, is the emission lifetime of trapped F molecules. Finally, the shortest component which is decaying in the highest energetic regions ($520 \pm 20$ nm) but rising (negative amplitude) in the lowest ones ($623 \pm 25$ nm) reflects a FRET from trapped F to RB dyes within the nanosheets.

As confirmation of the energy-transfer mechanism, we physically combined F@Z7-NS with RB@Z7-NS in a 1:1 and 10:1 ratios (Fig. 6a). The 1:1 ratio powder mixture shows near-yellow emission and, fascinatingly, the emission spectrum aligns with the mathematical superposition of the individual emission spectra of F@Z7-NS and RB@Z7-NS (Fig. 6a). In contrast, the emission spectrum for the dual-guest system shows far greater peak overlap, with a shift of the F emission band to longer wavelengths, while the RB emission shifts to shorter ones.

## Tuning the Dual-Guest System

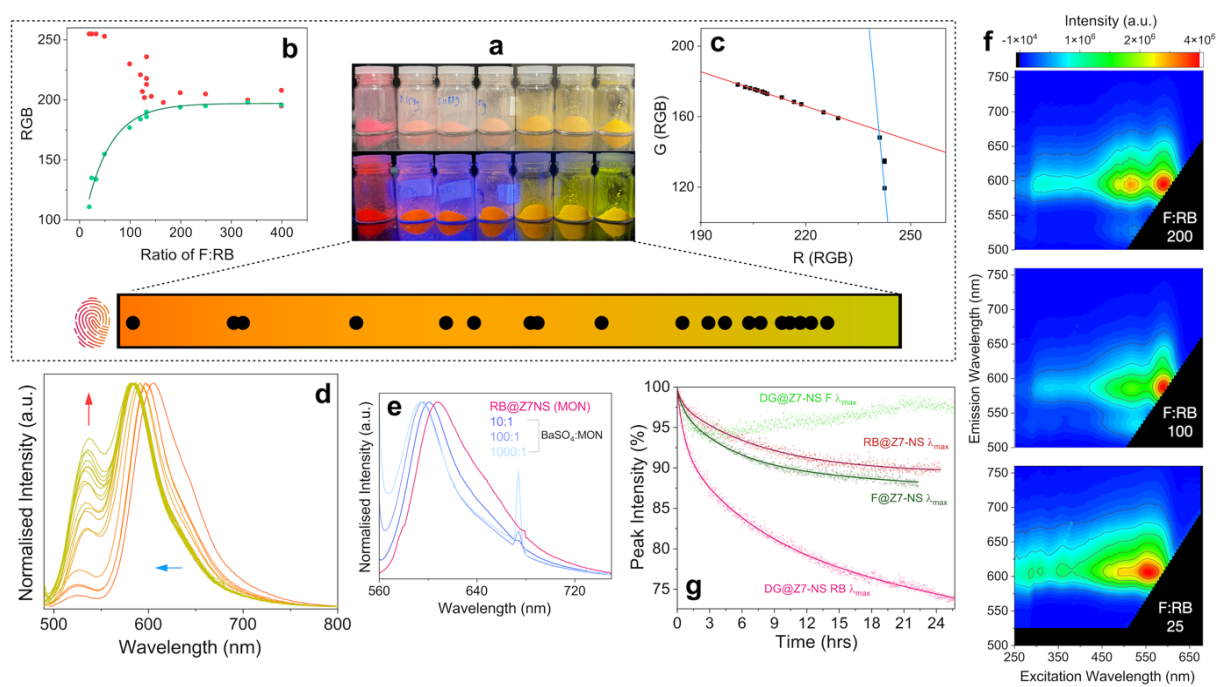

**Figure 7.** (a) Above: selection of synthesised samples in ambient conditions (top) and under UV (bottom), below: predicted finite spectrum of emission chromaticity for DG@Z7-NS derived from correlations, the 'emission chromaticity fingerprint', with emission chromaticity of synthesised samples indicated along the spectrum as black dots. (b) Correlation between synthesis ratio of F:RB and RGB of emission chromaticity. (c) R value of DG@Z7-NS materials correlated with G value (of RGB coordinates, derived from the emission spectra). (d) Emission spectra of various DG@Z7-NS samples excited at 470 nm; colours of each spectrum line indicate that sample's emission chromaticity RGB value. (e) Emission spectra of RB@Z7-NS when mixed with various ratios of $BaSO_4$. (f) Emission-excitation maps of three DG@Z7-NS samples at an F:RB ratio of 25, 100 and 200 (bottom to top). (g) Photostability of various guest@Z7-NS samples as a percentage of maximum absorption peak intensity over 24 hours with continuous exposure to respective sample $Abs_{max}$.

By varying the concentration of F during synthesis of the samples (Supplementary Table 5 and Fig. 28), a subset of twenty samples of DG@Z7-NS were obtained with distinct emission chromaticity coordinates (Fig. 7a, Supplementary Fig. 29). Significantly, the chromaticity of the system can be accurately quantified – allowing for precise fine-tuning of the emission colour of DG@Z7-NS. Upon plotting the synthesis ratio of F:RB against the RGB (red-green-blue) values of each sample (Fig. 7b), a good logarithmic correlation ($R^2 = 0.997$) was obtained between the F:RB concentration and the G value. Then, when comparing the G value to R value (Fig. 7c), we see two clear linear correlations with distinct domains ($R^2 = 0.996$). Together, these data can be used to identify the final emission colour of a dual-guest Z7-NS sample based simply on the ratio of F:RB used in the synthetic procedure. If a particular colour from green to red is desired, the ratio can be computed using Eqns.

(1) and (2). It is remarkable and of greatest significance is that this relationship confirms that there is only a discrete subset of colours (a 'emission chromaticity fingerprint') attainable by this dual-guest system: those that lie on the two G:R correlations (Fig. 7c), seen in the spectrum bar of Fig. 7a.

$$\boldsymbol{Eq.\,1}: Green\ value\ (G) = \frac{80.69604}{(1 + 1.29\mathrm{E}^{-4} \times Synthesis\ Ratio\ of\ F{:}RB)^{155.76324}}$$

$$\boldsymbol{Eq.\,2}: Red\ value\ (R) = \begin{cases} \dfrac{G - 5721}{-22}, & G < 170 \\[2mm] \dfrac{G - 338.15548}{-0.70008}, & G \geq 170 \end{cases}$$

**Equations 1 and 2.** Quantitative determination of DG@Z7-NS chromaticity based on ratio of F:RB used in the material's synthesis. The empirical coefficients were derived from Fig. 7(b) and 7(f)(inset).

The photophysical properties of the dual-guest materials were explored to determine the causes driving these chromaticity findings (Fig. 7, Supplementary Fig. 30-32). The spectral changes across the emission spectra of all 20 samples (Fig. 7d) also exhibit two clear processes occurring when the concentration of F increases in the system, with high-energy bands in the excitation spectra vary in similar stages. In the initial phase of 25 to 125 F:RB ratios, there is a blue shift of the RB system along with suppression of a shoulder band identifiable at ~650 nm. This corresponds to the first steep linear correlation in the R:G chromaticity relationship, with a dramatic increase of F emission with minimal RB value increase. By systematic mixing of the single-guest RB@Z7-NS sample with $BaSO_4$ in increasing ratios as shown in Fig. 7e, we suggest that these changes to be a cause of steric hindrance in the system. As the ratio increased from 10:1 to 1000:1, we saw a blueshift of emission peak, and the suppression of a shoulder at ~650 nm. As more salt crystals were added to RB@Z7-NS, we reasoned that a greater disturbance develops between Z7-NS crystals or even within the layers of sheets in single crystals. This will interfere with long-range interactions that we commented on before and which exists in the RB-Z7 system, particularly the J-aggregates responsible for the shoulder at 650 nm. The result is a general destabilisation of the system, and therefore a blue-shifted emission. The same measurement done with F resulted in very minimal shift and spectral shape variance comparably (Supplementary Fig. 33).

The first process ceases once a ratio of approximately 125 is reached, after which the second process dominates: the growth and narrowing of an emission band from F. The change in spectral patterns is most visible when comparing three emission maps (Fig. 7f), measured at samples with F:RB ratios of 25, 100 and 200, where a sharpening of emission bands is seen and the growth of a second local maxima at the absorption intensity maxima of fluorescein. Indeed, after the isosbestic point at the R value of 242 in the G:R correlation (Fig. 7c) a more subtle linear process also controls this relationship. As shown by the minimal shift from 1000:1 to 100:1 of $BaSO_4$:RB@Z7-NS (Fig. 7e), there reaches an equilibrium where steric hindrance no longer affects the energetics. From here, the second process dominates, being the increased emission from fluorescein and the interaction between F and RB molecules as observed in the optimised warm yellow-emitting sample (DG@Z7-NS).

Together, these data reveal that the competitive emission pathways within DG@Z7-NS dictate emission chromaticity. In further support of this are photostability measurements of DG@Z7-NS over 24 hours (Fig. 7g). Both guests decay at rates akin to the single-guest NS systems for the first 3 hours, but after the F band begins to intensify as RB continues to diminish. We attribute this to the photodegradation of RB monomers resulting in the reduction of F-RB energy transfer systems. This allows for increased emission from F monomers alone without absorption by RB. This continues until 21.5 hours, after which both guests appear to decay at a consistent rate evinced by proportional peak intensity loss in emission spectra. Importantly, at all times, yellow emission chromaticity is maintained (Supplementary Fig. 34).

## Conclusions

This work has demonstrated the versatile concept for first intercalation of luminescent guests into a 2D MON system. The hybrid system is also the first *in situ* synthesis of Z7-NS using a facile salt-templating methodology. This bottom-up approach produces highly stable, homogenous nanosheets that dispersed without aggregation in solution, and maintain dimensions of a few monolayer thickness (*ca*. 2−3 nm) while extended 5×15 μm$^2$ in plane. The selection and trapping of two fluorophores resulted in two single-guest LMONs, both exhibiting strong fluorescence controllable by guest loading concentration that affects guest packing and aggregation. Importantly, the systems show long-term physicochemical stability with a high resistance to leaching. Ultimately, we showed that combining two guests can result in a warm yellow emitting dual-guest intercalated MON that maintains the structural stability and nanosheet morphology of the single-guest systems. The system allows for an enhanced dual-guest energy-transfer interaction and, most remarkably, emission chromaticity that could be quantified so that tuneability became a precisely predictable process. This realisation that emission chromaticity is limited to a discrete subset of RGB values presents a strong theoretical foundation for rationally designing future LMON systems. By modelling predicted energetics of the dyes, followed by mapping a sample synthesis, it is possible to design systems of discrete colour ranges, with *a priori* 'emission chromaticity fingerprints', ideal for a range of applications from sensing to lighting, product quality assurance and security indicators.

## Methods

### Synthesis of ZIF-7-III Nanosheets (Z7-NS)

59.5 mg of $Zn(NO_3)_2$ (0.2 mmol) was dissolved in 2 mL of *N,N'*-dimethylformamide (DMF). The solution was then added to a conical flask containing 20 g of NaCl powder under vigorous magnetic stirring. 283.5 mg (2.4 mmol) of benzimidazole (bIm) was then dissolved in 2 mL of DMF. The solution was added dropwise to the conical flask followed by vigorous mixing for 12 hours. 250 mL of deionised $H_2O$ was then added to the flask and heated to 100 °C while stirring. The mixture was then siphoned into 50 mL centrifuge tubes and centrifuged at 10,000 rpm for 25 minutes. The collected powder at the bottom of each tube was combined into one tube, followed by further washing and centrifuging cycles (3 × 45 mL $H_2O$ and 5 × 40 mL MeOH). The final product was dried in the tube for 1 hour at 90 °C in an oven (~70-80%).

### Synthesis of Guests@Z7-NS Materials

The synthesis above was repeated except that initially 0.003 mmol (or 0.0003 mmol or 0.03 mmol) of Fluorescein (F) or Rhodamine B (RB) were solubilised with 2 mL of MeOH *via* sonication. This solution was mixed with the bIm solution before then adding to the $Zn(NO_3)_2$@NaCl vial.

**Powder X-Ray Diffraction (PXRD)**

PXRD patterns were collected using a Rigaku MiniFlex diffractometer equipped with a Cu Kα source and step size of 0.0025° at a scan rate of 0.04°/min. Samples were prepared using a 0.1 mm shallow well sample holder. High-resolution XRD patterns were collected at the Diamond Light Source I11 beamline using the MAC detection system. Calibrations were performed using Si powder standard (NIST SRM640c), with an X-ray beam energy of 15 keV (0.826834(10) Å) and scan time of 3600 seconds. Samples were prepared in borosilicate glass capillaries of 0.5 mm diameter. *d*-spacing was calculated using Bragg's Law.

**Attenuated Total Reflectance Fourier Transform Infrared Spectroscopy (ATR-FTIR)**

ATR-FTIR measurements of bulk Z7 and Z7-NS materials were performed using a Nicolet iS10 FTIR spectrometer. High-resolution synchrotron radiation infrared (SR-IR) measurements of the mid-IR and far-IR (FIR) spectra were performed at the Diamond Light Source B22 MIRIAM beamline. FIR and MIR spectra were collected under vacuum, using the ATR module installed on the Bruker Vertex 80V FTIR bench equipped with the DLaTGS detector. For improved signal-to-noise ratio, liquid helium-cooled bolometer was used in FIR measurements.

**Nano-FTIR**

Near-field optical measurements were collected using the neaSNOM instrument (neaspec GmbH) based on a tapping-mode AFM. The platinum-coated tip (NanoAndMore GmbH, cantilever resonance frequency 250 kHz and nominal tip radius ~20 nm) was illuminated by a Toptica broadband mid-infrared (MIR) femtosecond laser. Individual crystals of Z7 and Z7-NS type materials were analysed by averaging 4 measurements with 18 individual point spectra each. For each sample, at least 10 unique crystals or regions were probed. Each spectrum was acquired from an average of 20 Fourier-processed interferograms with 10 cm$^{-1}$ spectral resolution, 2048 points per interferogram, and an 18-ms integration time. The sample spectrum was normalised to a reference spectrum measured on a silicon surface to reconstruct the final nanoFTIR amplitude and phase. The continuous broadband MIR spectra were attained by combining two illumination sources, then the obtained spectra were combined at 1500 cm$^{-1}$. All measurements were carried out under ambient conditions (~40% RH).

**Thermogravimetric Analysis (TGA)**

TGA was performed using a TA Instruments Q50 TGA machine equipped with a platinum sample holder under an $N_2$ inert atmosphere at a heating rate of 10 °C/min from 30 to 800 °C.

**Raman Spectroscopy**

Raman spectra were collected using the Bruker MultiRAM Raman spectrometer with sample compartment D418, equipped with a Nd-YAG-Laser (1064 nm) and a LN-Ge diode as a detector. The laser power used for sample excitation was 50 mW, and 64 scans were accumulated at a resolution of 1 cm$^{-1}$.

**Diffuse Reflected Spectroscopy (DRS)**

The absorption spectra for samples were obtained using the 2600 UV-Vis spectrophotometer (Shimadzu) in the wavelength range of 200-1400 nm, equipped with an integrating sphere. The diffused reflectance spectra (DRS) were measured and converted using the Kubelka-Munk (KM) transformation to estimate the optical band gaps.

**Atomic Force Microscopy (AFM)**

AFM imaging was performed with a neaSNOM instrument (neaspec GmbH) operating in tapping mode. Height topography images were collected using the Scout350 probe (NuNano), which has a nominal tip radius of 5 nm, a spring constant of 42 N/m and resonant frequency of 350 kHz.

**Scanning Electron Microscopy (SEM)**

Backscattered electron and secondary electron SEM images were obtained at 10 keV under high vacuum using a SEM Tescan Lyra 3 (Tescan, Czech Republic) with secondary and backscattered electron imaging (SEI and BSE respectively) using a voltage from 10 keV to 15 keV. Samples were coated with gold (Au) with a thickness of 2.5 nm using SC7620 sputter coater (Quorum Technologies).

**Spectrofluorimetric Measurements**

Steady-state fluorescence spectra, steady-state diffuse reflectance spectra, photoluminescence quantum yield (PLQY), and TCSPC emission decay data were recorded using the FS-5 spectrofluorometer (Edinburgh Instruments) equipped with the appropriate modules for each specific experiment. For TCSPC measurements, the samples were pumped with a 365 nm EPLED picosecond pulsed laser source. Lifetime fitting of the time constants from decay data was performed using the Fluoracle software. Excitation spectra exhibit artefacts (sharp peaks) from equipment operation that are not removed in presented data.

**Fluorescence Lifetime Imaging (FLIM)**

Fluorescence lifetime images (FLIM) were recorded using an inverted-type scanning confocal fluorescence microscope (MicroTime-200, Picoquant, Berlin) with a 60× NA1.2 Olympus water immersion objective, and a 2D piezo scanner (Physik Instrumente). A 470-nm pulsed diode laser (pulse width ~40 ps) was employed as the excitation source. A dichroic mirror (AHF, Z375RDC), a 500-nm long-pass filter (AHF, HQ500lp), a 100-μm pinhole, and an avalanche photodiode detector (MPD, PDM series) were used to collect the emission. Moreover, the FLIM and emission decays of the DG@Z7-NS sample were collected using two band pass filters (520 ± 20 nm and 623 ± 25 nm) to selectively record the emission region of F and RB, respectively. The emission spectrum was recorded using a spectrograph (Andor SR 303i-B) equipped with a 1600×200 pixels EMCCD detector (Andor Newton DU-970N-BV) coupled to the Micro-Time-200 system.

**Transmission Electron Microscopy (TEM)**

TEM characterisation was performed at 200 keV accelerating voltage using a $LaB_6$ electron source using a JEM-2100 electron microscope. The images were recorded on a Gatan Orius camera through Digital Micrograph.

**Nuclear Magnetic Resonance (NMR) Guest Loading Analysis**

Samples for NMR were dissolved in a solution composed of 500 μL methanol-d4 and 50 μL DCl / $D_2O$ (35 wt.%). All NMR spectroscopy was done at 298 K using a Bruker AvanceIII spectrometer operating at 600 MHz, equipped with a BBO cryoprobe. Data were collected using a relaxation delay of 20 s, with 128k points and a sweep width of 19.8 ppm, giving a digital resolution of 0.18 Hz. Data was processed using Bruker

Topspin with a line broadening of 1 Hz and 2 rounds of zero-filling. The loading amount was calculated from the molar ratio of compound to bIm. In order to calculate the molar ratio, peaks corresponding to each compound were integrated and normalised according to the number of protons giving rise to the signal. For fluorescein (F), the doublet at approximately 8.30 ppm was used, which corresponds to a single proton (that in the ortho position relative to the carboxyl group). For Rhodamine B (RB) the broad multiplet at approximately 8.22 ppm was used, which corresponds to the equivalent single proton in the ortho position relative to the carboxyl group. For benzimidazole the singlet at approximately 9.33 ppm was used, which corresponds to the single proton of the imidazole group, adjacent to the two nitrogens. Global spectral deconvolution (in the MestReNova software package) was used to pick and integrate the peaks.

## Acknowledgements

D.A.S. acknowledges the scholarships from the General Sir John Monash Foundation and the Clarendon Fund. J.C.T. and S.M. thank the ERC Consolidator Grant (PROMOFS 771575) and EPSRC (EP/R511742/1) for funding the research. M.G. and A.D. were supported by grants PID2020-116519RB-I00 funded by MCIN/AEI/10.13039/501100011033 and by the EU; SBPLY/19/180501/000212 and SBPLY/21/180501/000108 funded by JCCM and by the EU through "Fondo Europeo de Desarollo Regional" (FEDER). We acknowledge the Diamond Light Source for the provision of beamtime SM27504 at B22 MIRIAM *via* Drs. Mark Frogley and Gianfelice Cinque, and rapid access beamtime CY29415 at I11 *via* Dr. Sarah Day. We would like to acknowledge Dr. Cyril Besnard and Professor Alexander Korsunsky for the acquisition of the FESEM images. We thank the Research Complex at Harwell (RCaH) for access to materials characterisation facilities.

## Author contributions

Conceptualisation was done by D.S and J.-C.T. Methodology was developed by D.S. Synthesis was performed by D.S. AFM, PXRD, nano-FTIR, FIR, MIR and FS-5 spectrofluorimetric measurements were performed by D.A.S. SEM characterisation was performed by C.B. and D.S. TEM characterisation was performed by I.G. with analysis of the data by D.A.S. FLIM experiments were performed and analysed by M.G. and A.D. NMR was performed by N.A. TGA, band gap and Raman data was collected by S.M, analysis performed by D.A.S. Manuscript was drafted by D.A.S, with review and editing by D.A.S., M.G., A.D. and J.-C.T. Supervision by J.-C.T.

## Competing interests

The authors declare no competing interests.

## Additional information

Supplementary information. The online version contains supplementary materials available at {enter URL from publisher}.

# Supporting Information

# Quantifiably Tuneable Luminescence by Ultra-Thin Metal-Organic Nanosheets via Dual-Guest Energy Transfer

Dylan A. Sherman,[a] Mario Gutiérrez,[b] Ian Griffiths,[c] Samraj Mollick,[a] Nader Amin,[d] Abderrazzak Douhal,[b] and Jin-Chong Tan [a,*]

[a]Multifunctional Materials & Composites (MMC) Laboratory, Department of Engineering Science, University of Oxford, Parks Road, Oxford OX1, United Kingdom.
[b]Departamento de Química Física, Facultad de Ciencias Ambientales y Bioquímica, INAMOL, Universidad de Castilla-La Mancha, Toledo 45071, Spain.
[c]Department of Materials, University of Oxford, 16 Parks Road, Oxford OX1 3PH, United Kingdom.
[d]Department of Chemistry, University of Oxford, Mansfield Road, Oxford OX1 3TA, United Kingdom.

*Corresponding author: jin-chong.tan@eng.ox.ac.uk

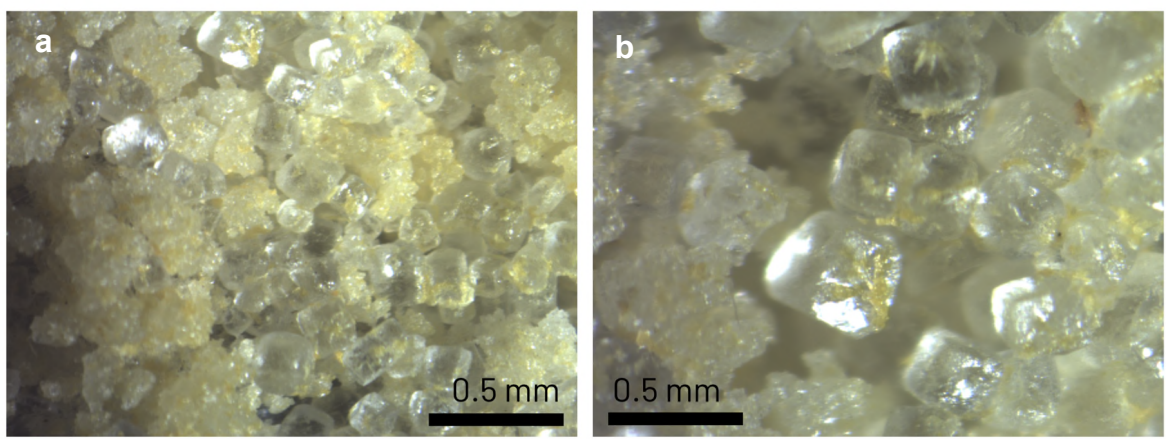

**Figure 1.** (a-b) Stereoscopic optical microscope images (x50) of NaCl crystals coated with F@Z7-NS growth prior to washing.

**a** F@Z7-NS 10⁻¹

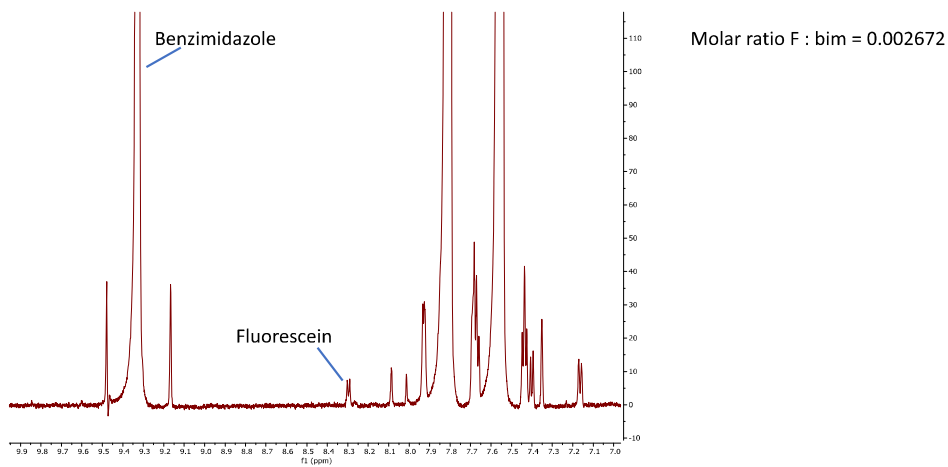

Benzimidazole

Fluorescein

Molar ratio F : bim = 0.002672

**b** F@Z7-NS 10⁻²

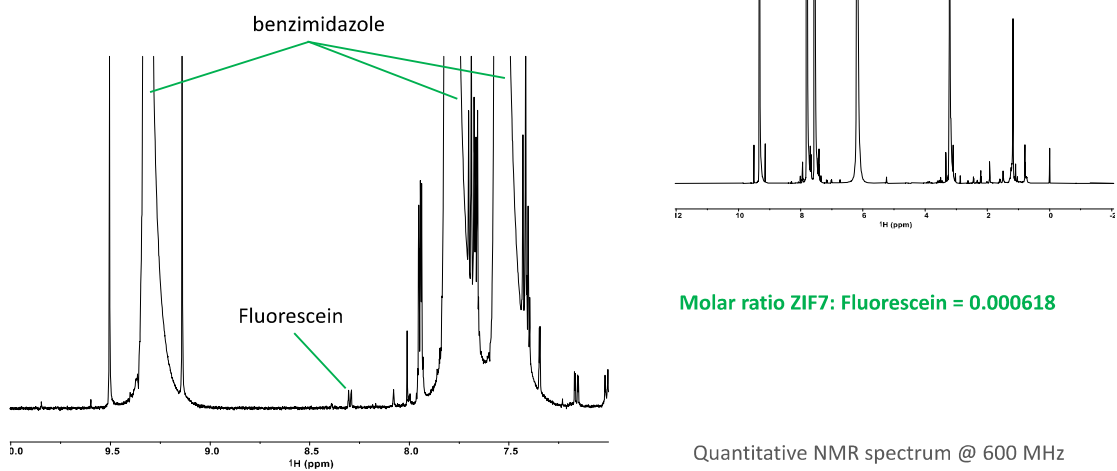

benzimidazole

Fluorescein

**Molar ratio ZIF7: Fluorescein = 0.000618**

Quantitative NMR spectrum @ 600 MHz

**c** F@Z7-NS 10⁻³

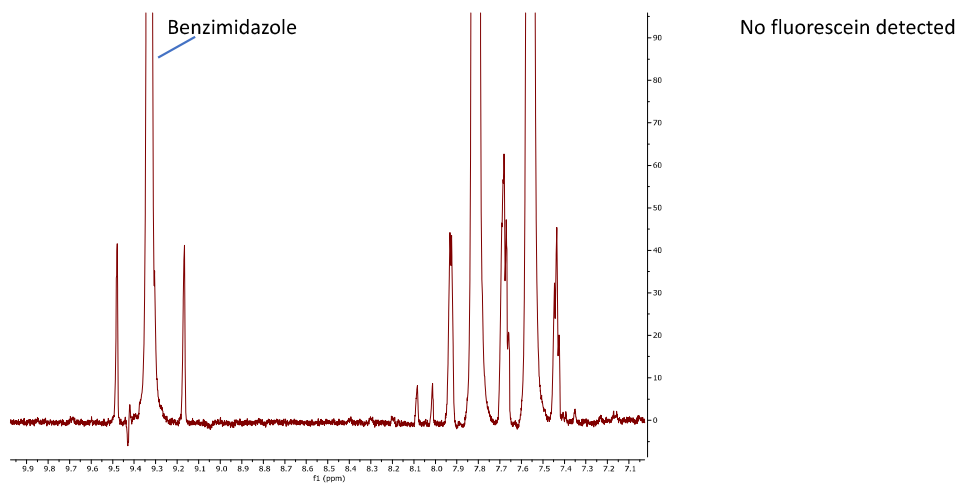

Benzimidazole

No fluorescein detected

**d**  RB@Z7-NS 10⁻¹

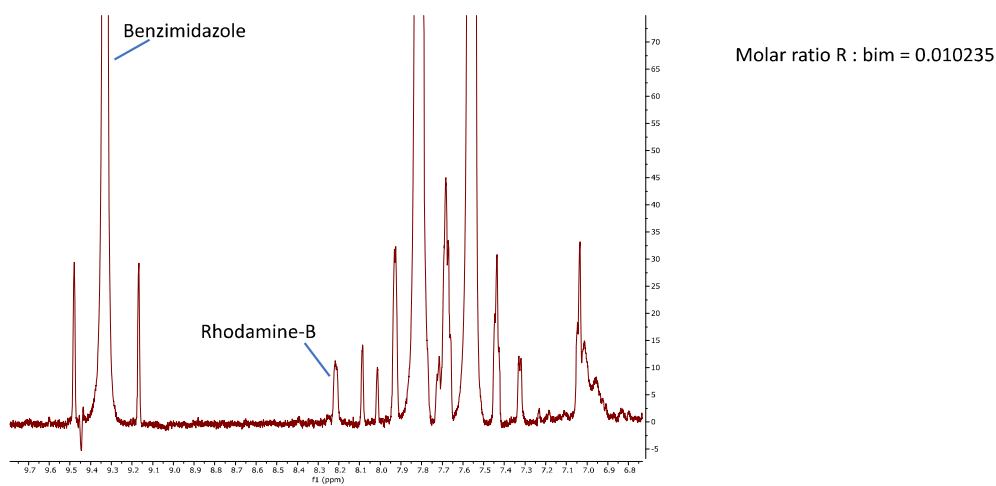

Molar ratio R : bim = 0.010235

**e**  RB@Z7-NS 10⁻²

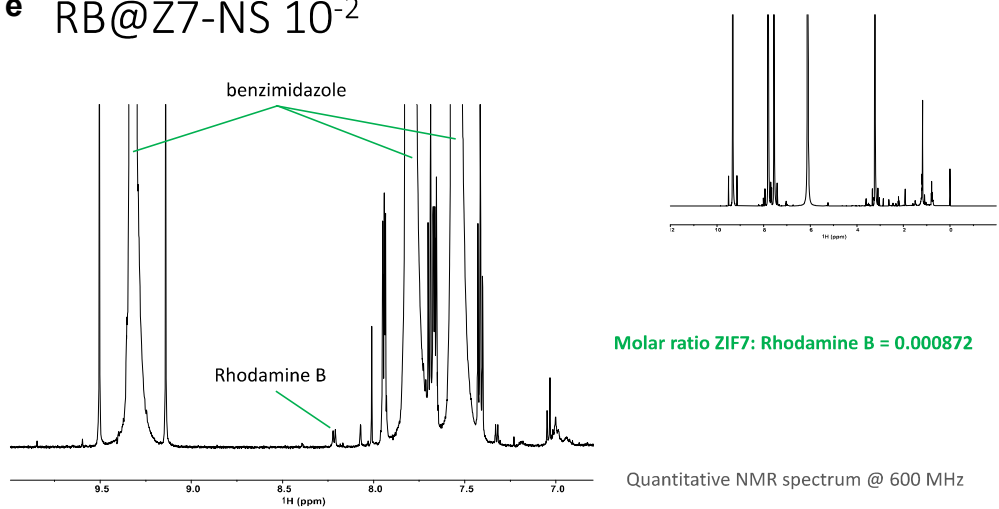

Molar ratio ZIF7: Rhodamine B = 0.000872

Quantitative NMR spectrum @ 600 MHz

**f**  RB@Z7-NS 10⁻³

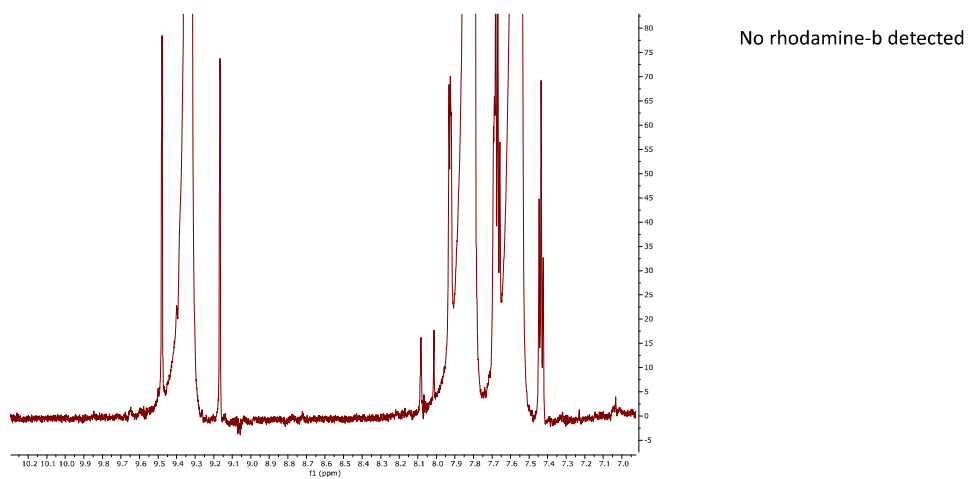

No rhodamine-b detected

**g** DG-Z7-NS (1)

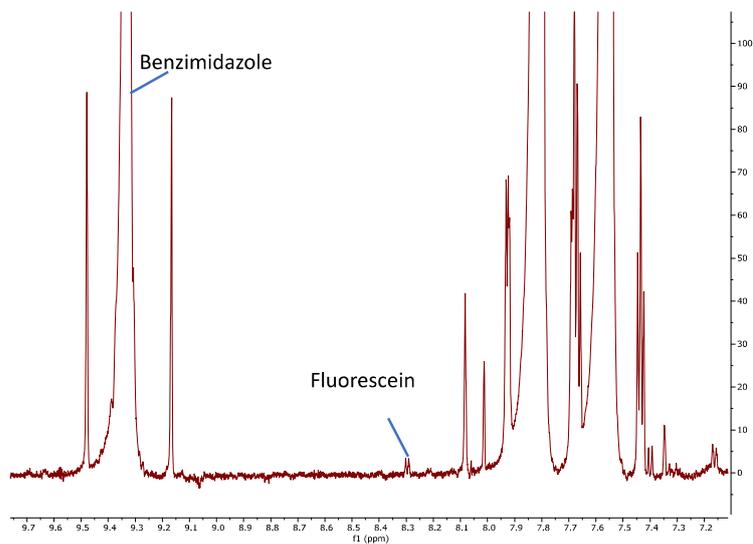

Molar ratio F : bim = 0.000444

**h** DG-Z7-NS (4)

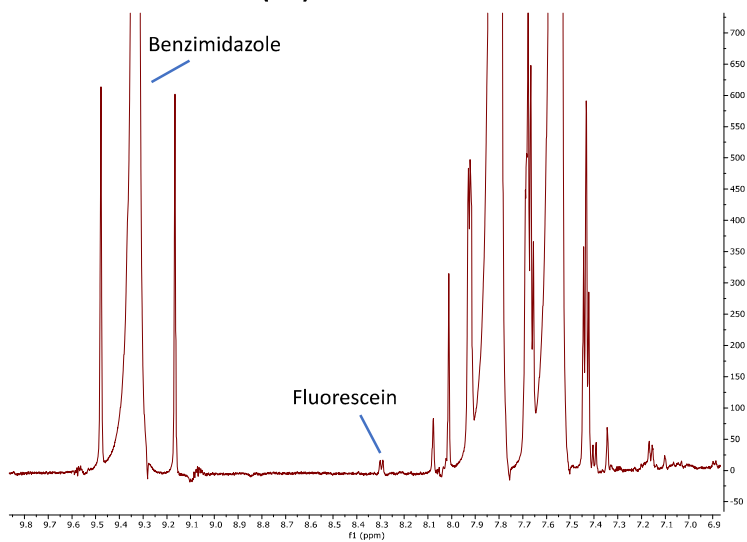

Molar ratio F : bim = 0.000453

**j** DG-Z7-NS (5)

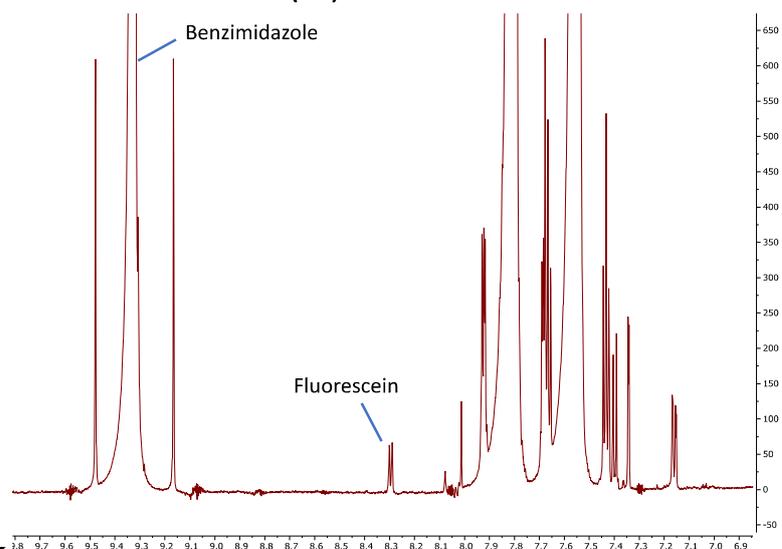

Molar ratio F : bim = 0.001688

**k** DG-Z7-NS (13)

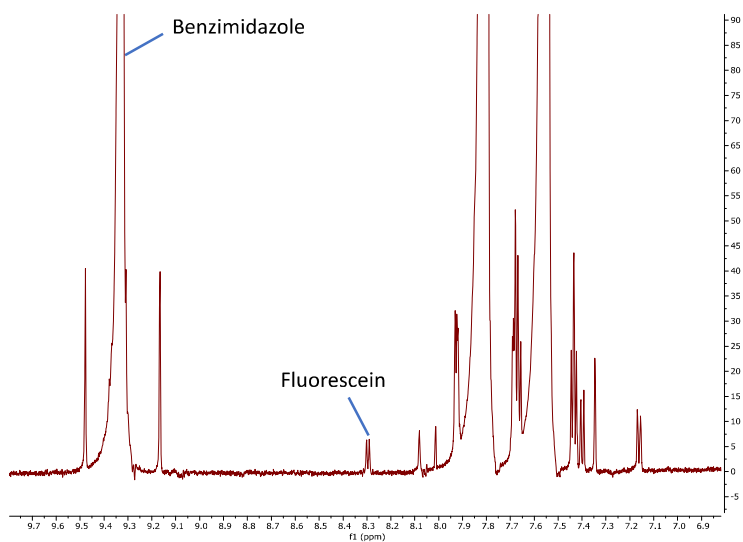

Molar ratio F : bim = 0.00209

**Figure 2.** (a-k) ¹H NMR Digest Spectra of Z7-NS Samples with calculated ratio of guest: ligand in each sample.

| Sample | Detected ratio of guest:blm in products via $^1$H NMR | Synthesis guest quantity (mmol) | Synthesis ratio of guest:blm |
|---|---|---|---|
| F@Z7-NS-10$^{-3}$ | *Below detection limit* | 0.0003 | 0.000125 |
| F@Z7-NS-10$^{-2}$ | 0.000618 | 0.003 | 0.00125 |
| F@Z7-NS-10$^{-1}$ | 0.002672 | 0.03 | 0.0125 |
| RB@Z7-NS-10$^{-3}$ | *Below detection limit* | 0.0003 | 0.000125 |
| RB@Z7-NS-10$^{-2}$ | 0.000872 | 0.003 | 0.00125 |
| RB@Z7-NS-10$^{-1}$ | 0.010235 | 0.03 | 0.0125 |
| DG@Z7-NS-1 (F content only) | 0.000444 | 0.003 | 0.00131 |
| DG@Z7-NS-4 (F content only) | 0.000453 | 0.003 | 0.00131 |
| DG@Z7-NS-5 (F content only) | 0.001688 | 0.003 | 0.0013 |
| DG@Z7-NS-13 (F content only) | 0.00209 | 0.003 | 0.0014 |

**Table 1.** Guest loadings of selected samples determined by $^1$H NMR.

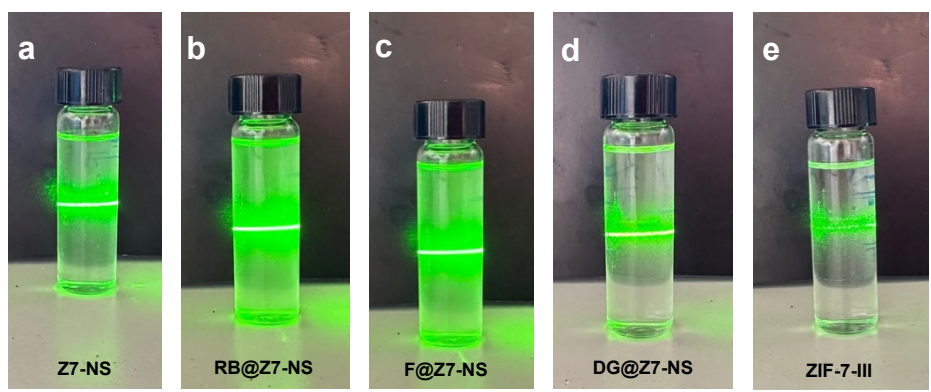

**Figure 2.** (a-d) Z7-NS particles in MeOH exhibiting Tyndall scattering effect resulting from uniform dispersions of nanoparticles. (e) ZIF-7-III particles MeOH exhibiting minimal Tyndall scattering due to aggregates.

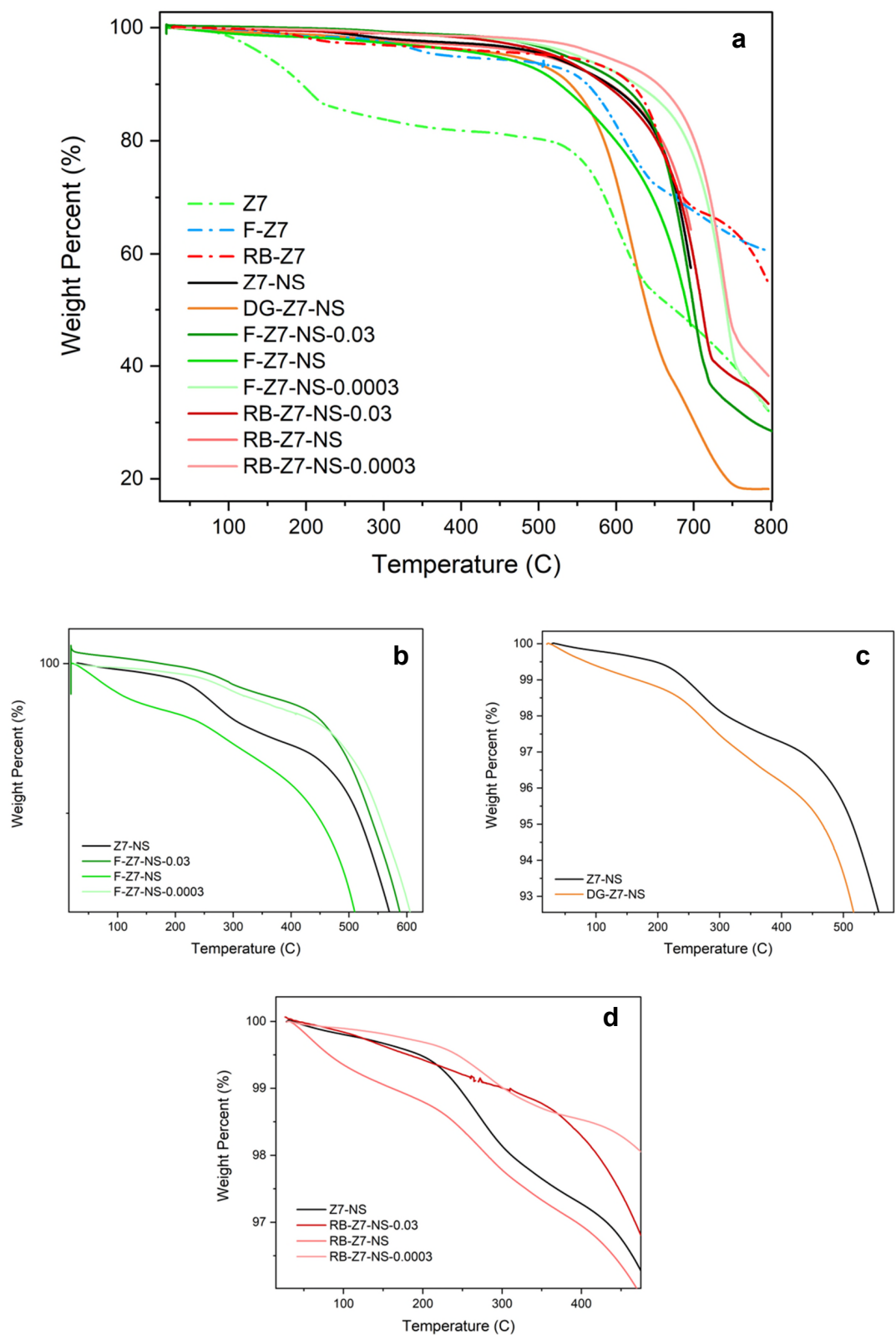

**Figure 4**. (a-d) TGA curves of synthesised Z7 and Z7-NS materials with various guests.

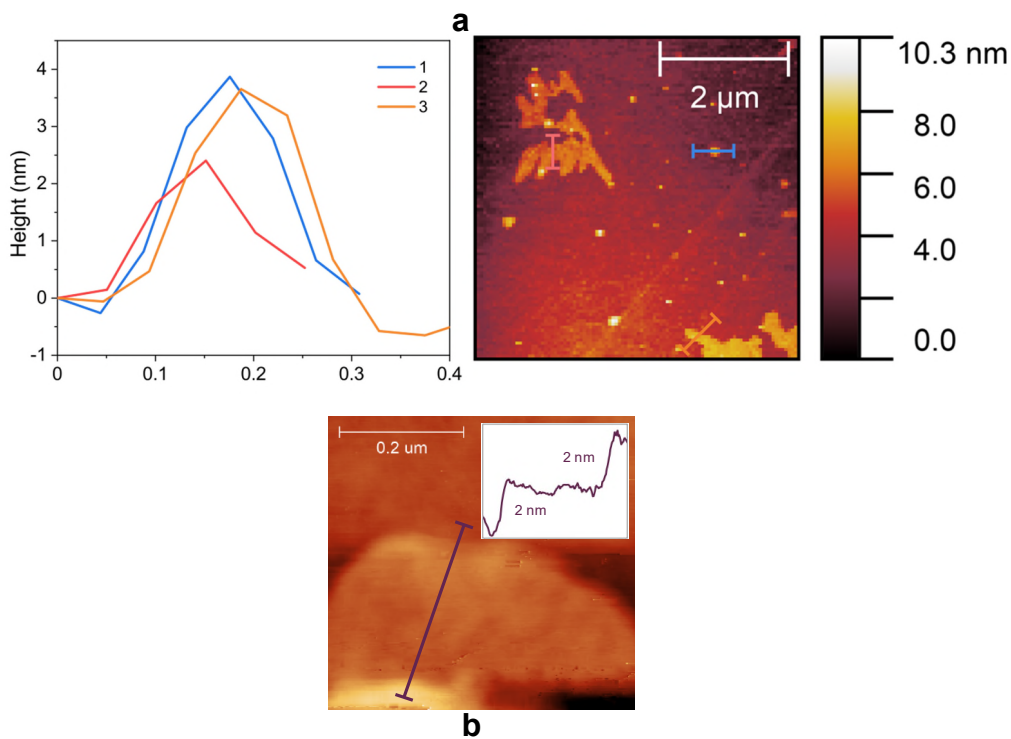

**Figure 5**. (a-b) AFM mapping of Z7-NS with annotated height profiles indicated by lines in corresponding colours.

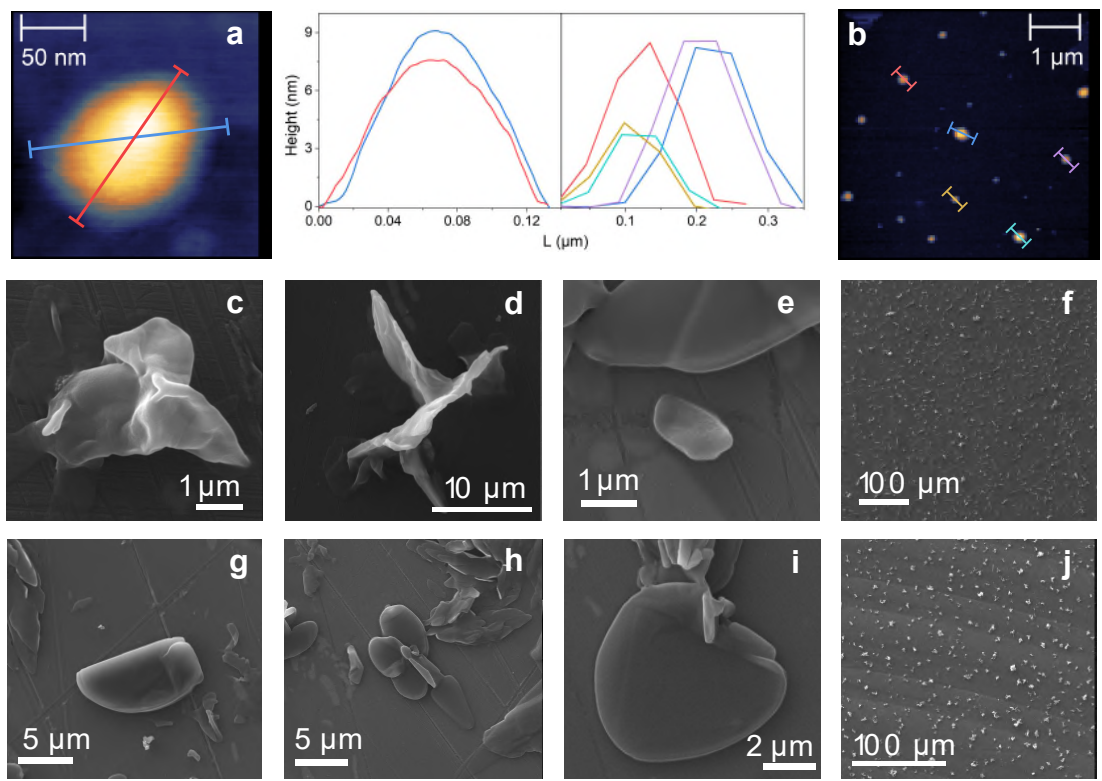

**Figure 6**. (a-b) AFM mapping and (c-j) SEM imaging of RB@Z7-NS samples.

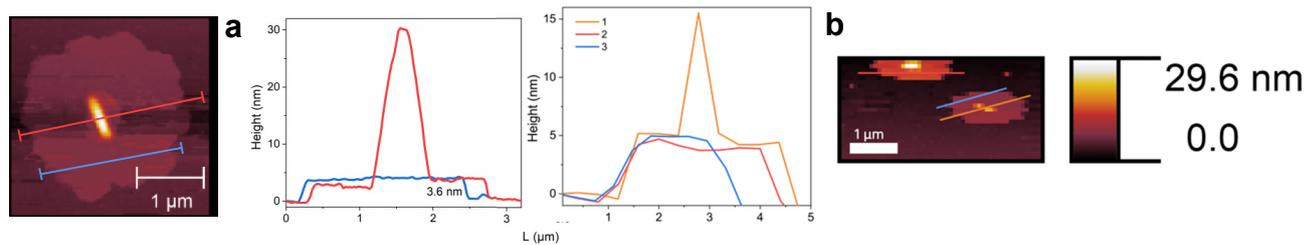

**Figure 7**. (a-b) AFM mapping of F@Z7-NS samples.

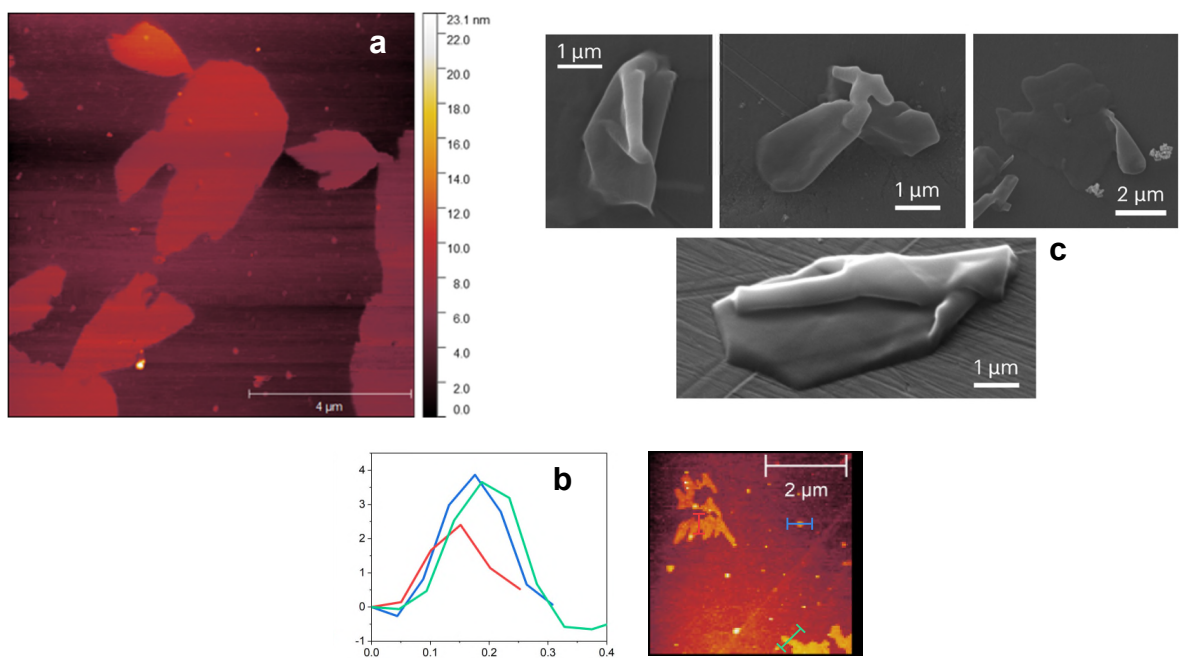

**Figure 8**. (a-b) AFM mapping and (c) SEM imaging of DG@Z7-NS samples.

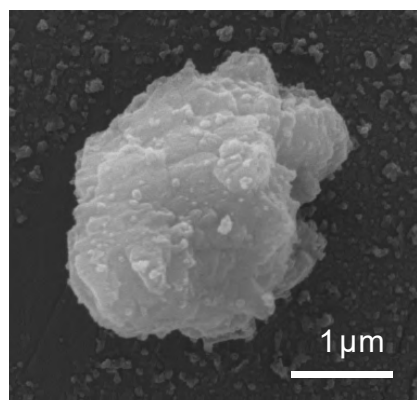

**Figure 9**. SEM image of ZIF-7-III layered particle.

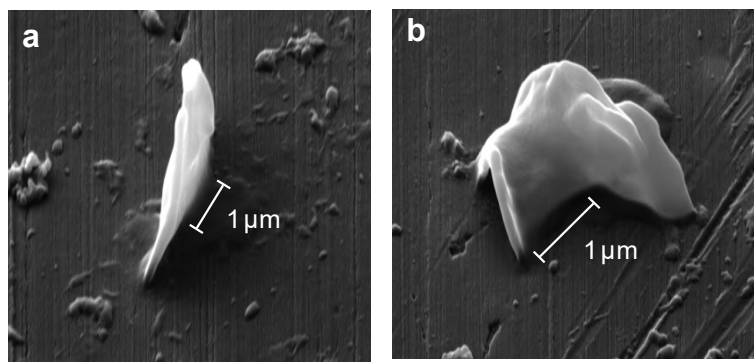

**Figure 10**. (a-b) SEM imaging of Z7-NS at a 45° tilt.

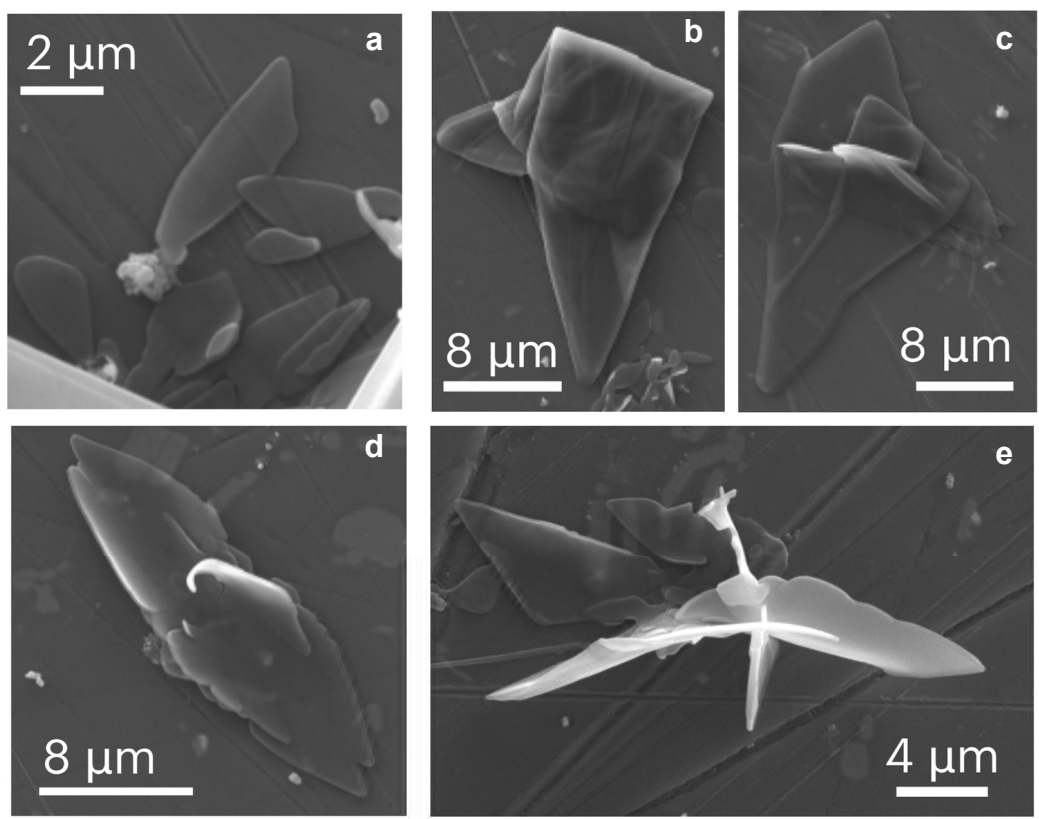

**Figure 11**. (a-e) SEM imaging of Z7-NS after soaking in MeOH for 6 months.

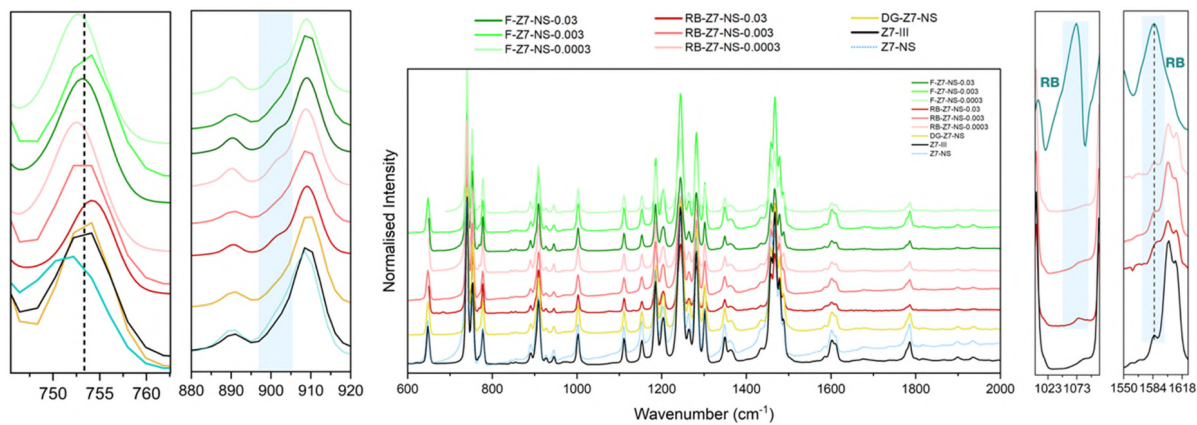

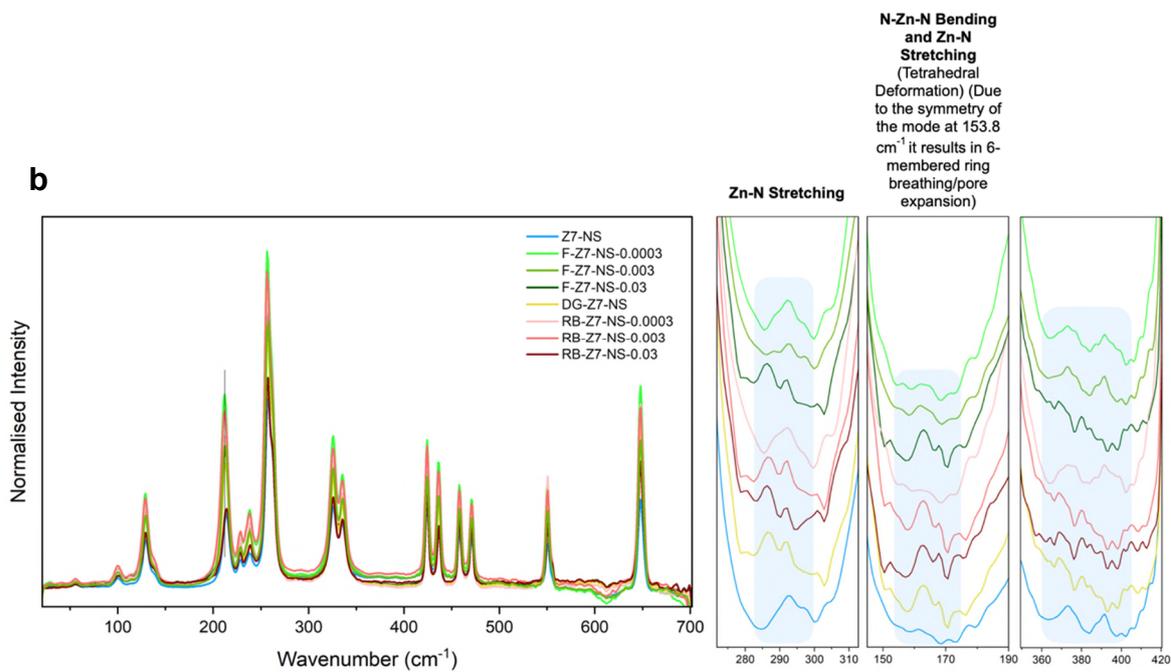

**Figure 12**. Synchrotron Radiation Infrared (SR-IR) Spectra in the (a) MID-IR and (b) FAR-IR of Z7-NS and guest@Z7-NS samples.

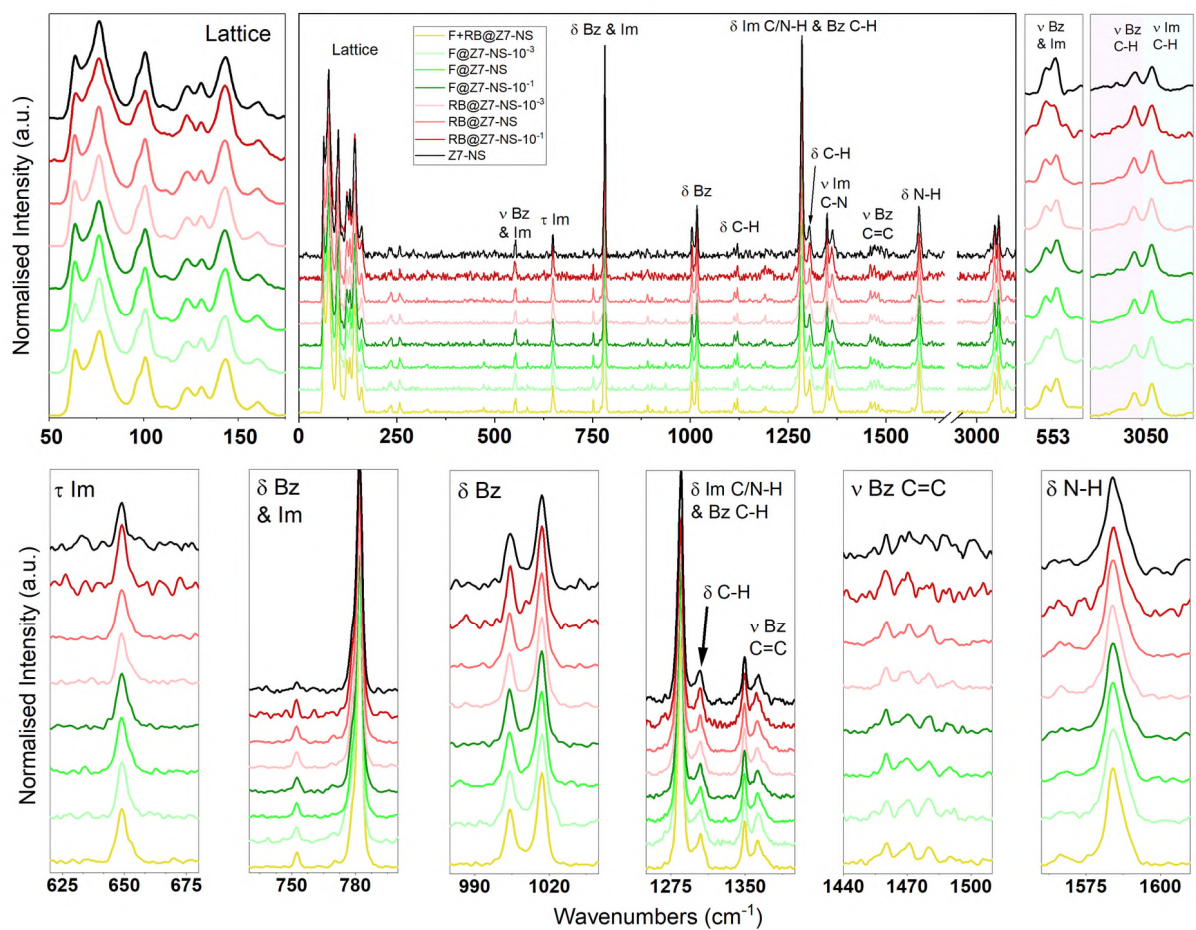

**Figure 13**. Raman spectra of Z7-NS and guests@Z7-NS samples. ν: stretching; δ: in-plane bending; τ: torsion. Bz: benzene ring; Im: imidazole ring (assignments based on Zhao, P.; Lampronti, G. I.; Lloyd, G. O.; Wharmby, M. T.; Facq, S.; Cheetham, A. K.; Redfern, S. A. Phase Transitions in Zeolitic Imidazolate Framework 7: The Importance of Framework Flexibility and Guest-Induced Instability. Chem. Mater. **26**, 1767−1769 (2014)).

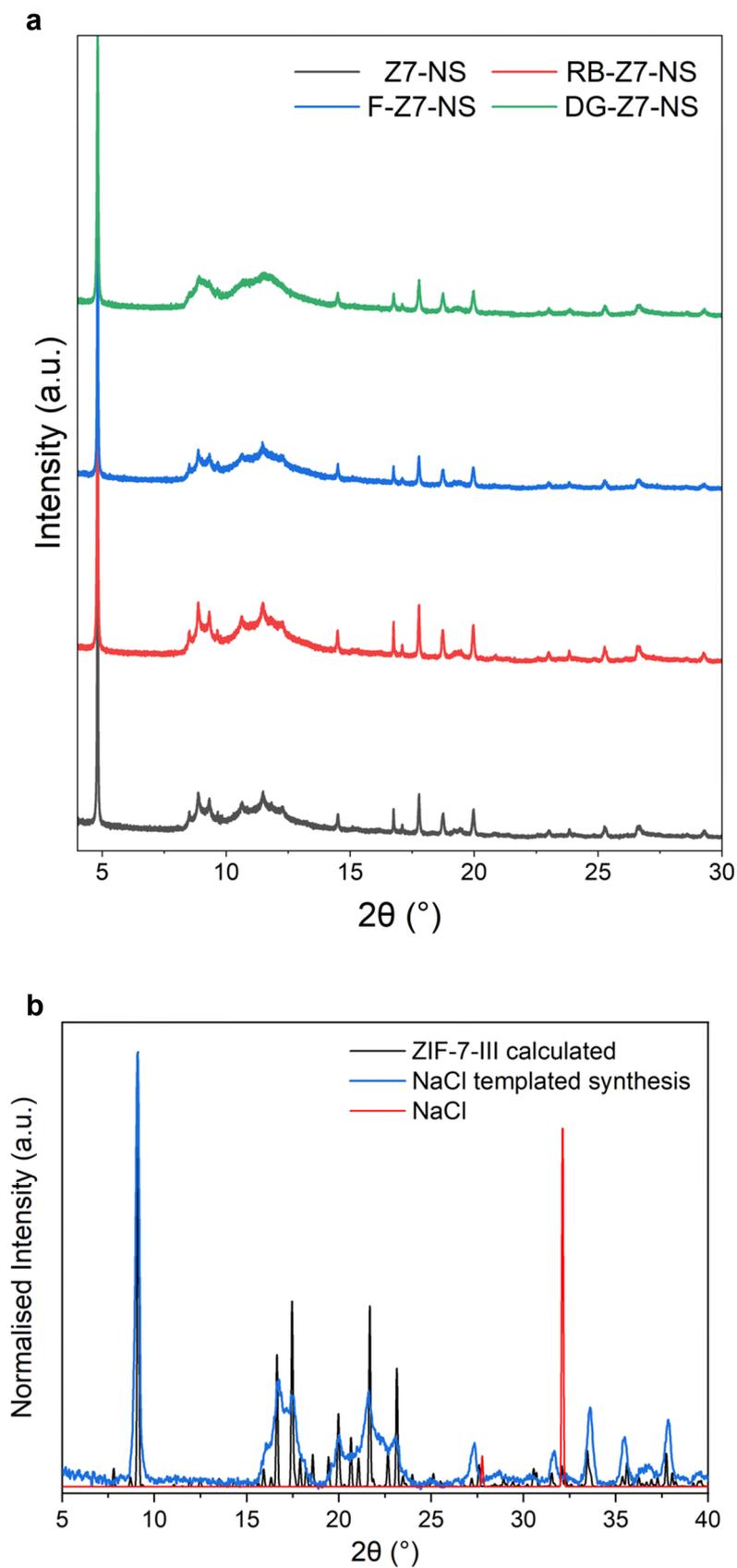

**Figure 14**. (a) Synchrotron PXRD spectra of Z7-NS samples and (b) PXRD of ZIF-7-III calculated compared with NaCl and Z7-NS salt templated synthesis product.

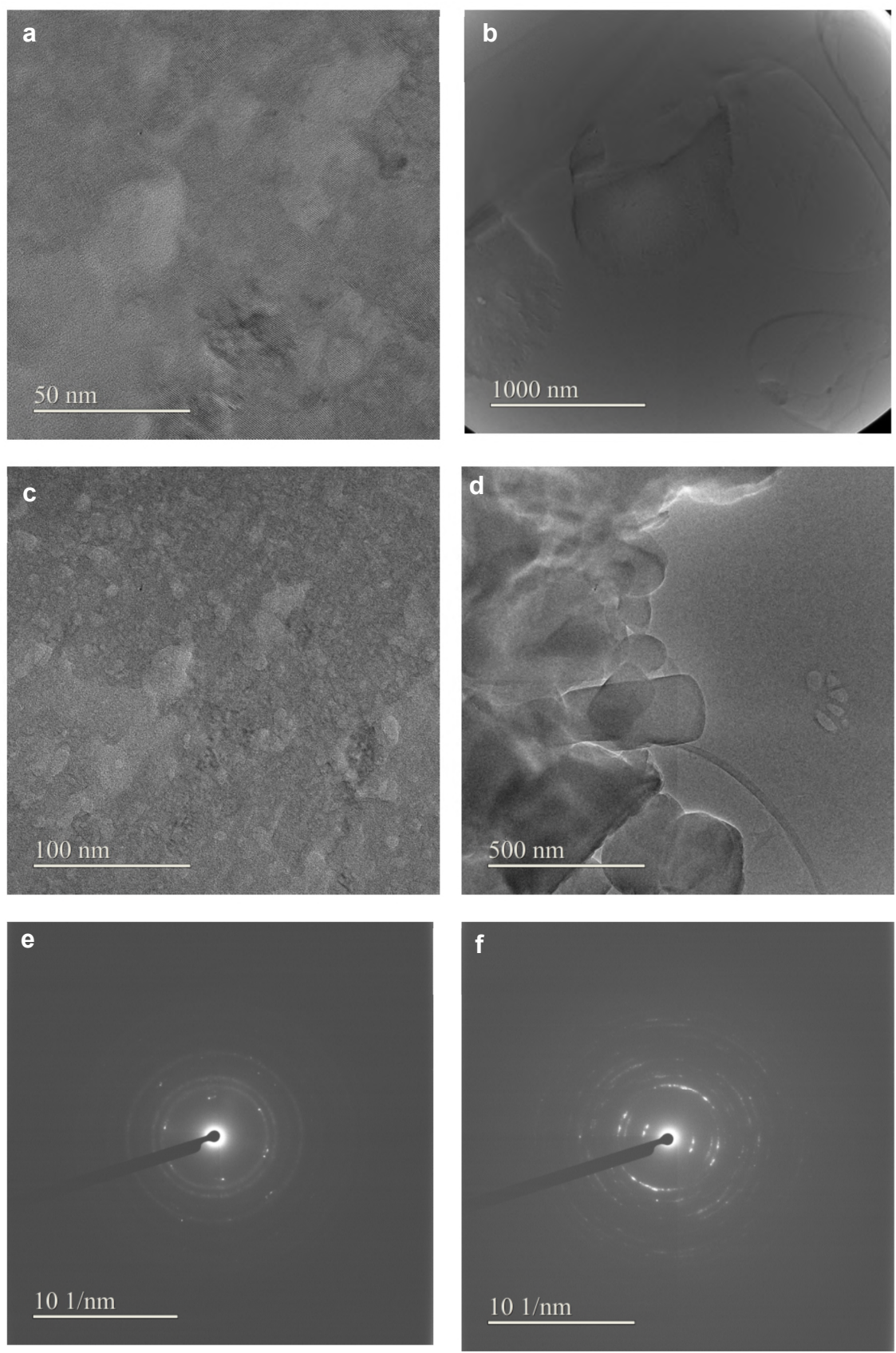

**Figure 15**. (a-d) HRTEM imaging of Z7-NS particles. (e-f) SAED patterns of the Z7-NS particle observed in (b).

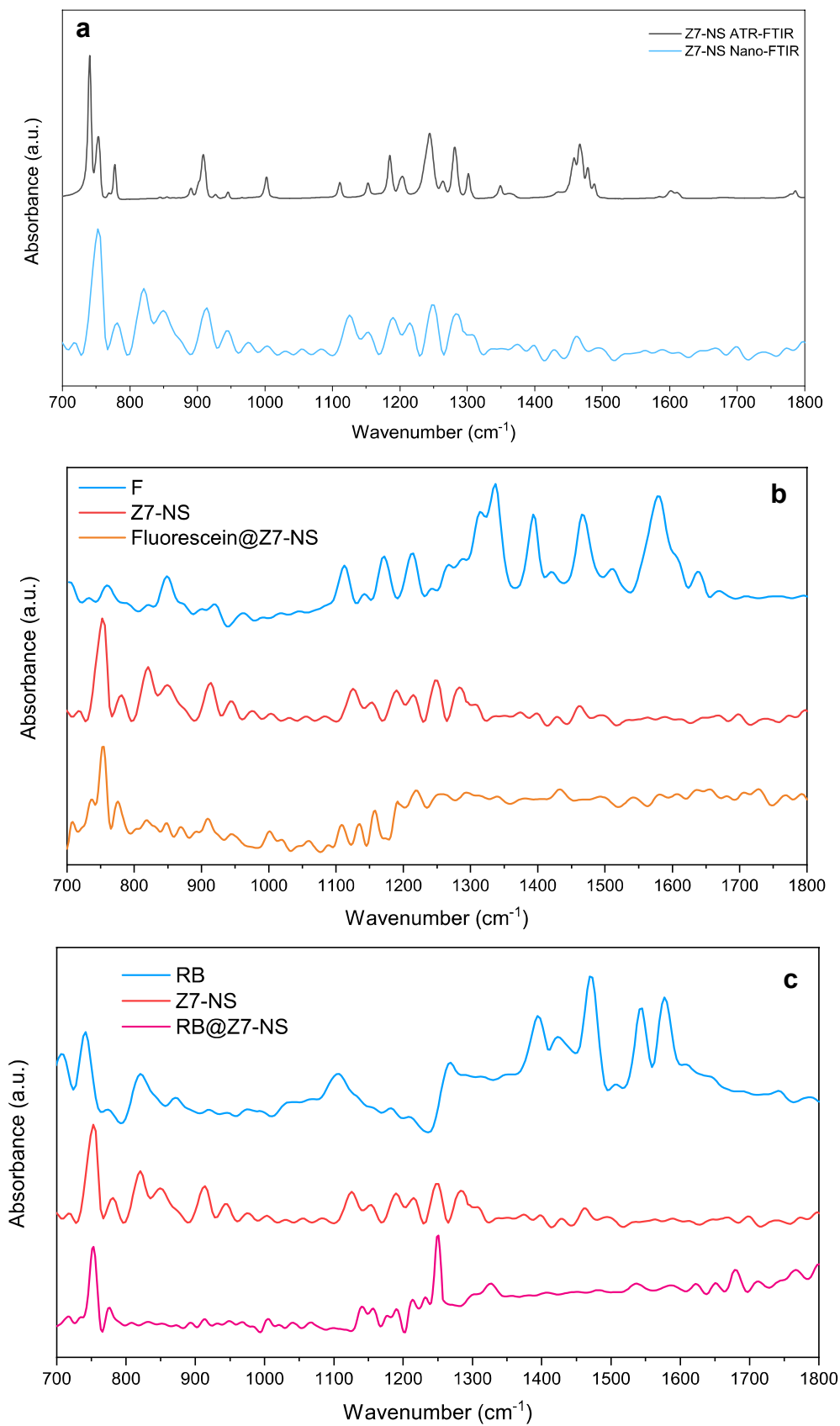

**Figure 16**. (a) Nano-FTIR of Z7-NS compared with ATR-FTIR, (b-c) Extended wavelength range nano-FTIR of RB@Z7-NS and F@Z7-NS compared with F and RB molecules and pure Z7-NS.

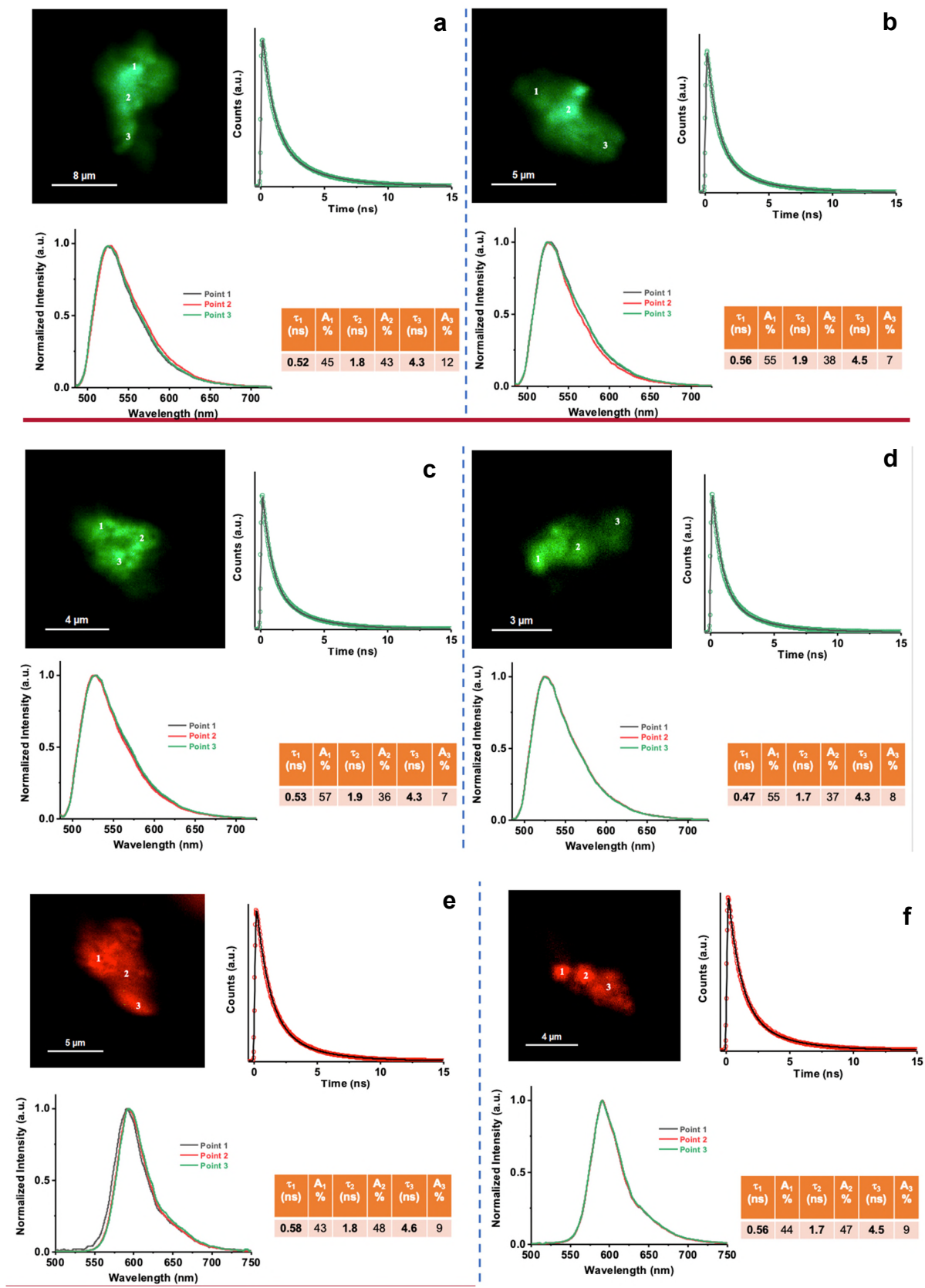

**Figure 17**. (a-d) F@Z7-NS and (e-f) RB@Z7-NS FLIM imaging, emission spectra and lifetime decay curves with estimated lifetime contributions.

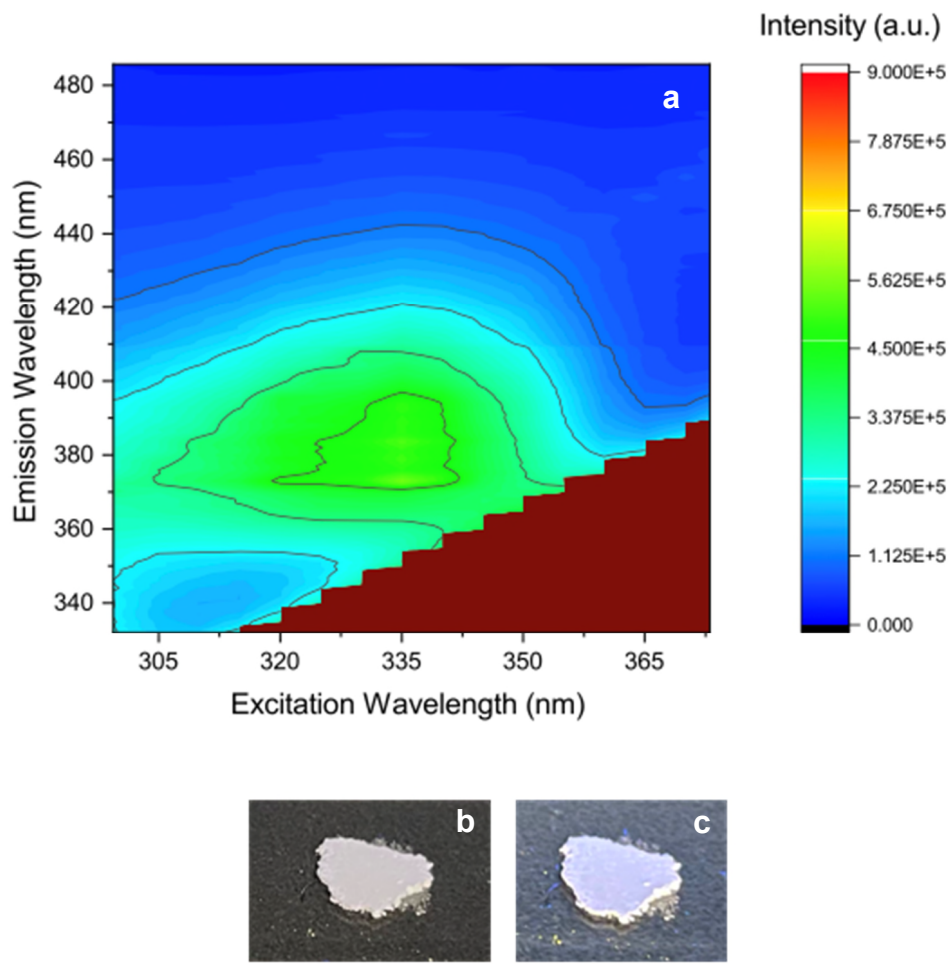

**Figure 18**. (a) Emission map of Z7-NS (pristine host framework). (b) Z7-NS sample in ambient conditions and (c) under UV indicating negligible sample emission.

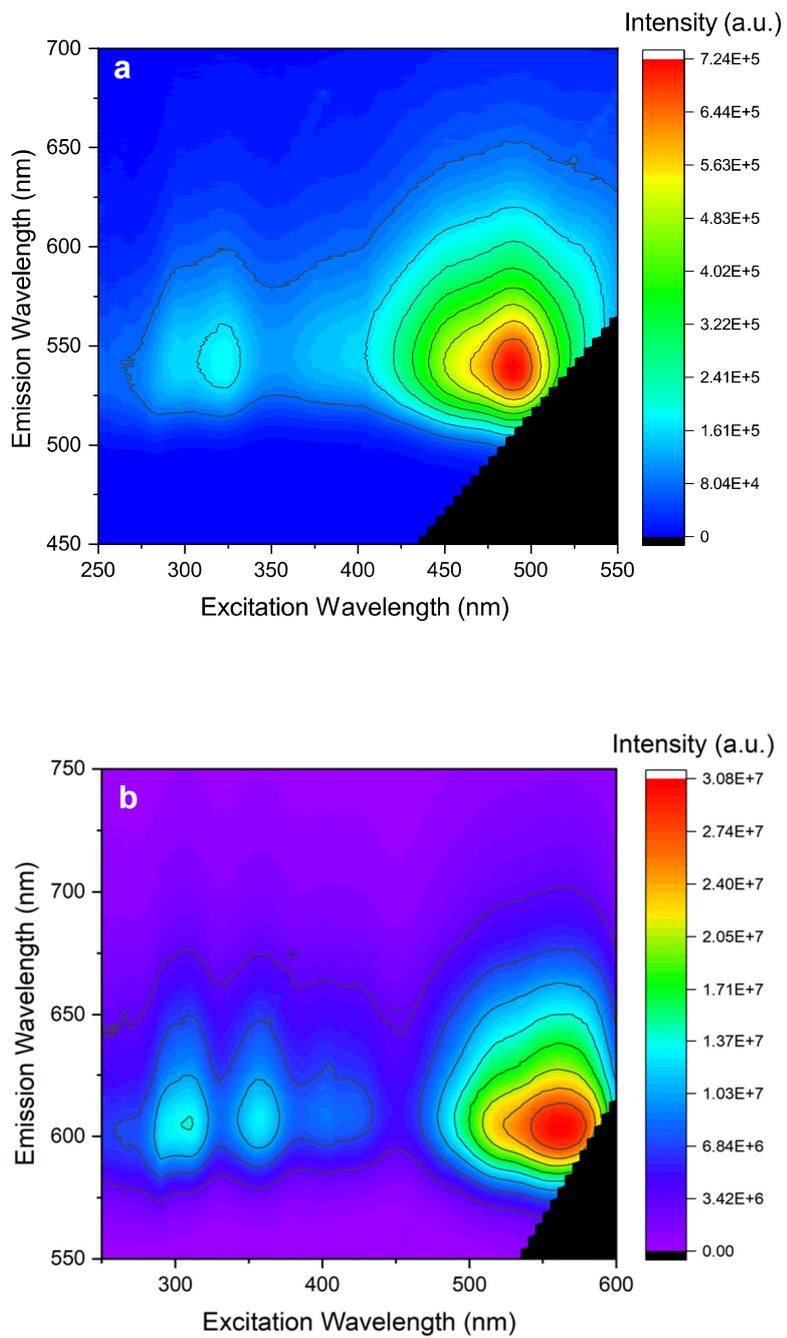

**Figure 19**. Fluorescence emission maps of (a) F@Z7-NS and (b) RB@Z7-NS.

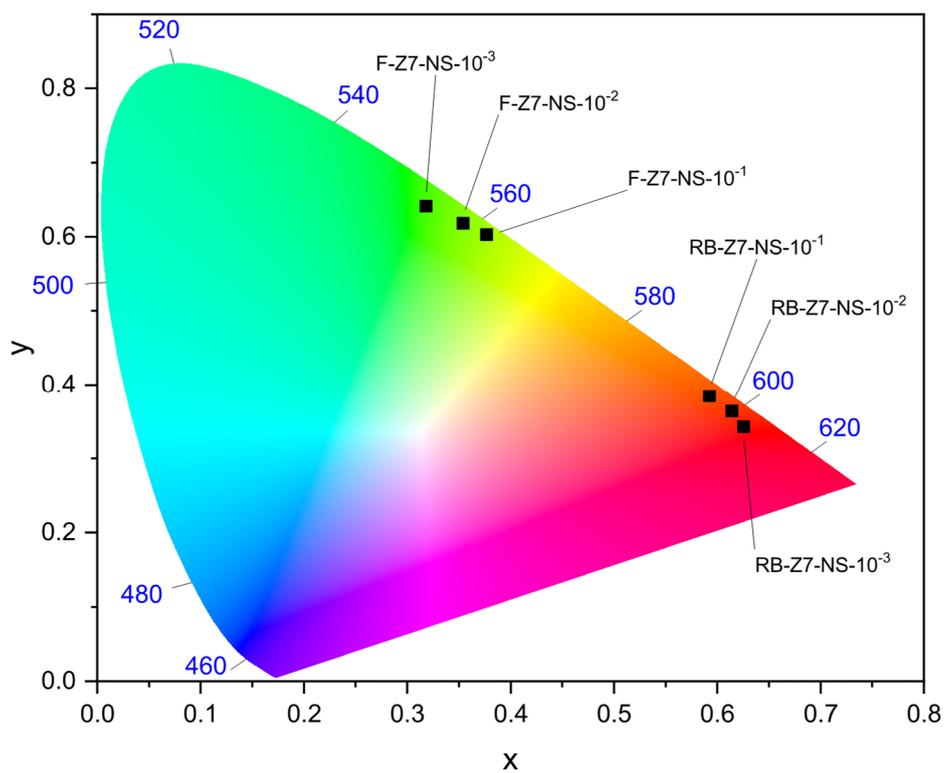

**Figure 20**. Emission chromaticity of F@Z7-NS and RB@Z7-NS samples dependent on guest loading.

| | $\lambda_{obs}$ [nm] | $\tau_1$[ns] | $a_1$ | $c_1$[%] | $\tau_2$[ns] | $a_2$ | $c_2$[%] | $\chi^2$ |
|---|---|---|---|---|---|---|---|---|
| **F@Z7-NS 10$^{-3}$** | 520 | 0.1166 | 0.1001 | 4.59 | 1.9539 | 0.0581 | 44.68 | 1.1514 |
| | 540 | 0.1906 | 0.0531 | 3.56 | 2.3427 | 0.0579 | 47.66 | 1.1294 |
| | 560 | 0.1649 | 0.0638 | 3.65 | 2.4255 | 0.0571 | 48.11 | 1.1579 |
| **F@Z7-NS 10$^{-2}$** | 520 | 0.1394 | 0.2144 | 6.82 | 1.114 | 0.0552 | 61.76 | 1.1961 |
| | 540 | 0.1614 | 0.1702 | 5.6 | 1.2269 | 0.0577 | 60.06 | 1.1735 |
| | 560 | 0.1951 | 0.1441 | 5.12 | 1.3292 | 0.0561 | 56.96 | 1.1548 |
| **F@Z7-NS 10$^{-1}$** | 520 | 0.1205 | 0.1964 | 20.91 | 1.2411 | 0.0571 | 52.1 | 1.2842 |
| | 540 | 0.1784 | 0.1352 | 19.35 | 1.4814 | 0.055 | 54.57 | 1.2858 |
| | 560 | 0.129 | 0.1645 | 19 | 1.3362 | 0.0583 | 55.47 | 1.1666 |
| **Fluorescein** | 515 | 4.0859 | 0.193 | 100 | | | | 1.331 |

**Table 2**. Values of time constants ($\tau_i$), normalised pre-exponential factors ($a_i$), and fractional contributions ($c_i = \tau_i \cdot a_i$) of the emission decay of F@Z7-NS samples with different guest loading in solid state excitation at 362.5 nm.

| | $\lambda_{obs}$ [nm] | $\tau_1$[ns] | $a_1$ | $c_1$[%] | $\tau_2$[ns] | $a_2$ | $c_2$[%] | $\tau_3$[ns] | $a_3$ | $c_3$[%] | $\chi^2$ |
|---|---|---|---|---|---|---|---|---|---|---|---|
| **RB@Z7-NS $10^{-3}$** | 600 | 0.1794 | 0.0333 | 1.87 | 2.4486 | 0.0544 | 41.85 | 4.9006 | 0.0366 | 56.28 | 1.2539 |
| | 620 | 0.348 | 0.0225 | 2.3 | 2.8778 | 0.0551 | 46.5 | 5.3999 | 0.0324 | 51.2 | 1.1492 |
| | 640 | 0.2308 | 0.0457 | 3.06 | 3.1514 | 0.0609 | 55.7 | 5.8273 | 0.0244 | 41.24 | 1.0435 |
| **RB@Z7-NS $10^{-2}$** | 600 | 0.6659 | 0.0148 | 3.04 | 2.7434 | 0.0671 | 57.06 | 5.693 | 0.0226 | 39.9 | 1.1909 |
| | 620 | 0.9477 | 0.0115 | 3.12 | 2.9755 | 0.0652 | 55.73 | 5.9025 | 0.0243 | 41.15 | 1.1567 |
| | 640 | 1.088 | 0.0156 | 4.76 | 3.3165 | 0.0689 | 64.17 | 6.5396 | 0.0169 | 31.07 | 1.2355 |
| **RB@Z7-NS $10^{-1}$** | 600 | 0.2262 | 0.0704 | 7.19 | 1.6895 | 0.0697 | 53.09 | 4.4253 | 0.0199 | 39.72 | 1.2026 |
| | 620 | 0.3962 | 0.0607 | 10.3 | 2.1456 | 0.0659 | 60.57 | 5.5904 | 0.0122 | 29.13 | 1.2685 |
| | 640 | 0.4477 | 0.0597 | 11.61 | 2.2448 | 0.0646 | 63.03 | 5.8304 | 0.01 | 25.36 | 1.302 |
| **Rhodamine B** | 571 | 2.4137 | 0.215 | 100 | | | | | | | 1.315 |

**Table 3**. Values of time constants ($\tau_i$), normalised pre-exponential factors ($a_i$), and fractional contributions ($c_i = \tau_i \cdot a_i$) of the emission decay of RB@Z7-NS samples with different guest loading in solid state excitation at 362.5 nm.

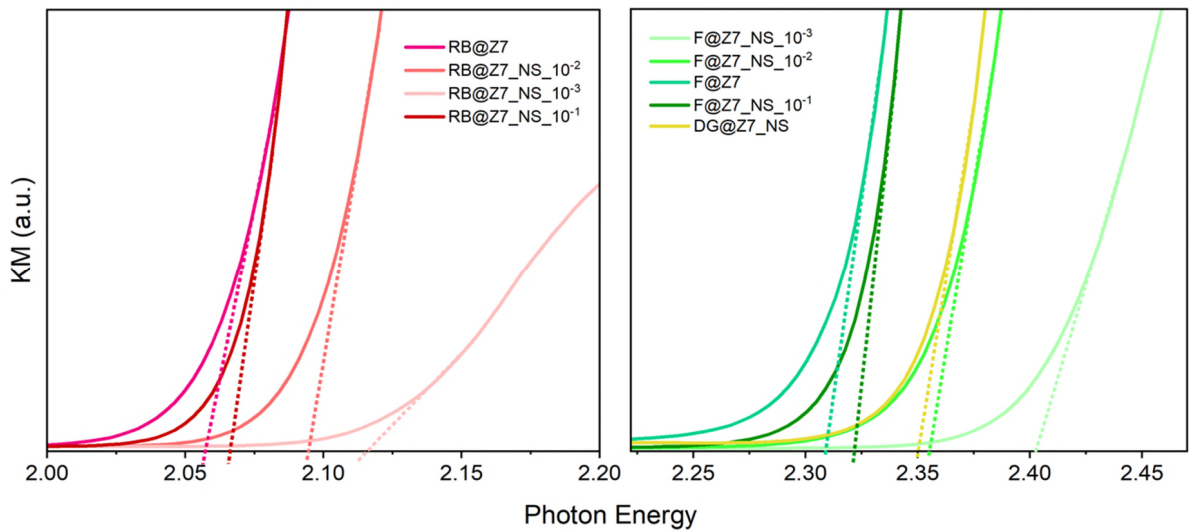

**Figure 21**. Band gaps of DG@Z7-NS, F@Z7-NS and RB@Z7-NS materials compared with F and RZ encapsulated in ZIF-7 (3D) framework.

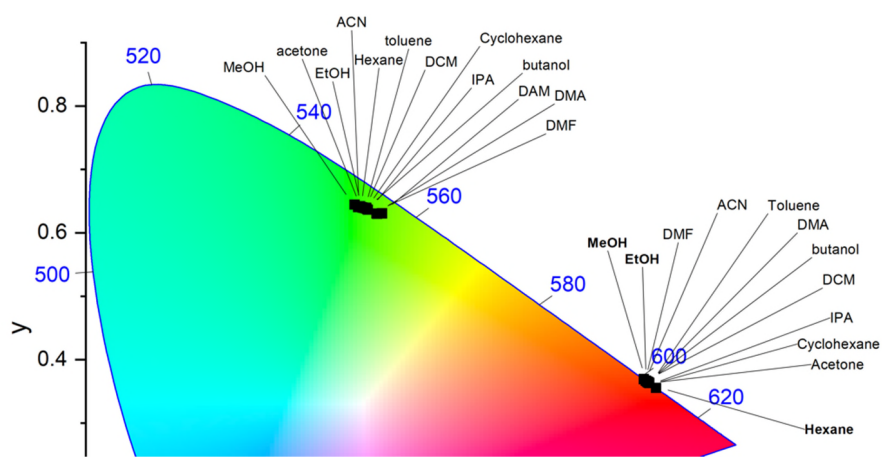

**Figure 22**. Solvatochromic response of F@Z7-NS and RB@Z7-NS in various organic solvents

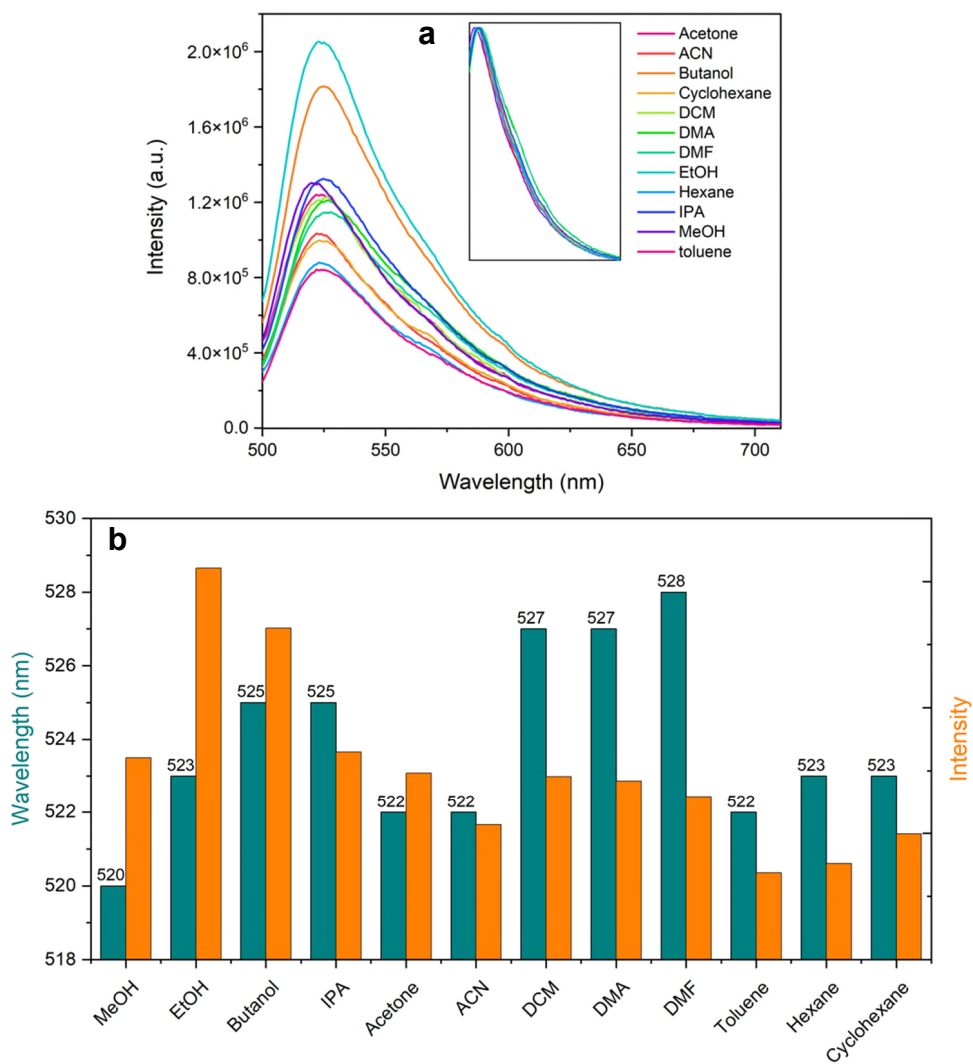

**Figure 23**. (a) Solvatochromic shift in emission spectra of F@Z7-NS in various organic solvents. (b) comparison of Em$_{max}$ peak intensity and wavelength of F@Z7-NS in varied organic solvents.

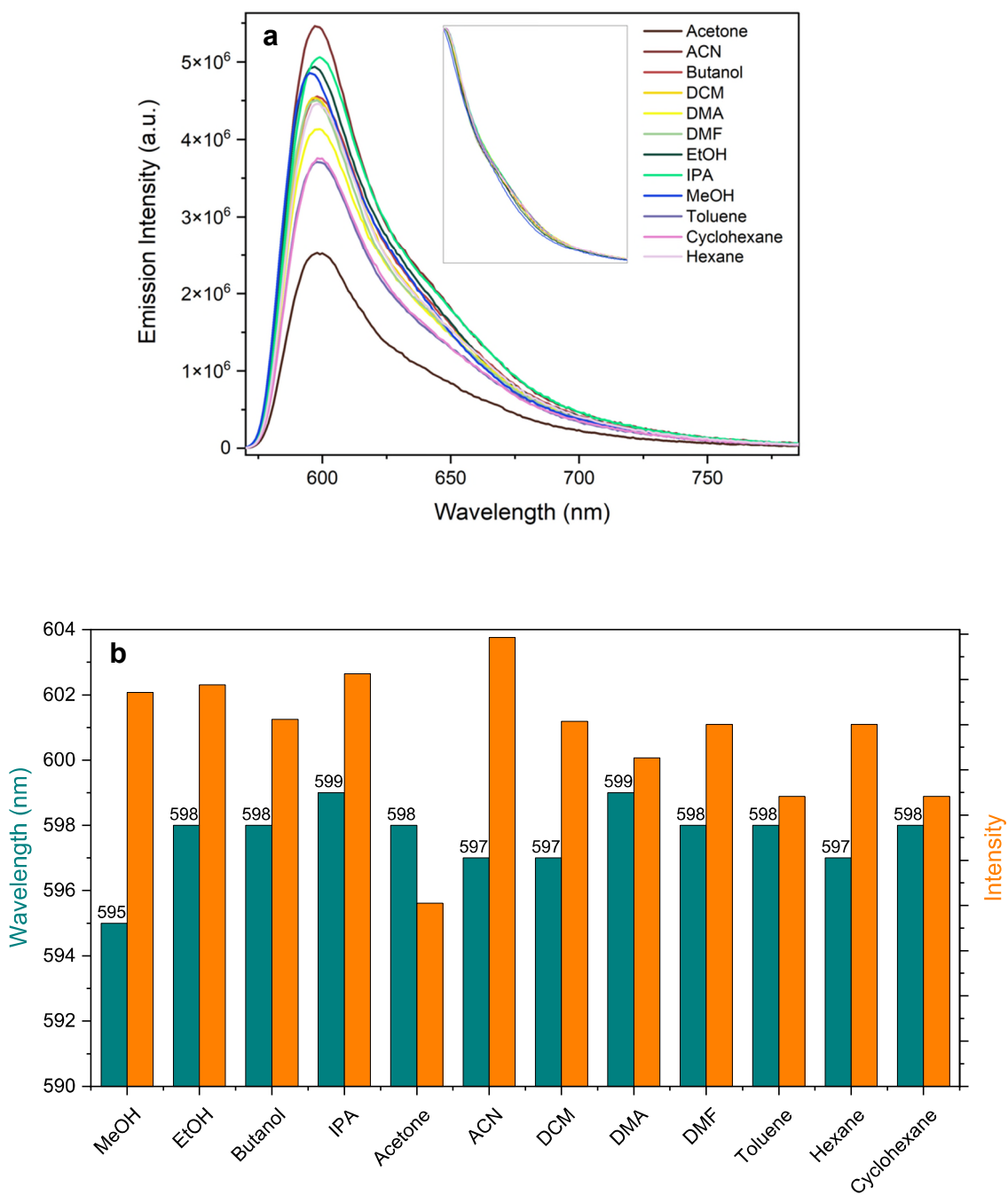

**Figure 24**. (a) Solvatochromic shift in emission spectra of RB@Z7-NS in various organic solvents. (b) comparison of Em$_{max}$ peak intensity and wavelength of RB@Z7-NS in varied organic solvents.

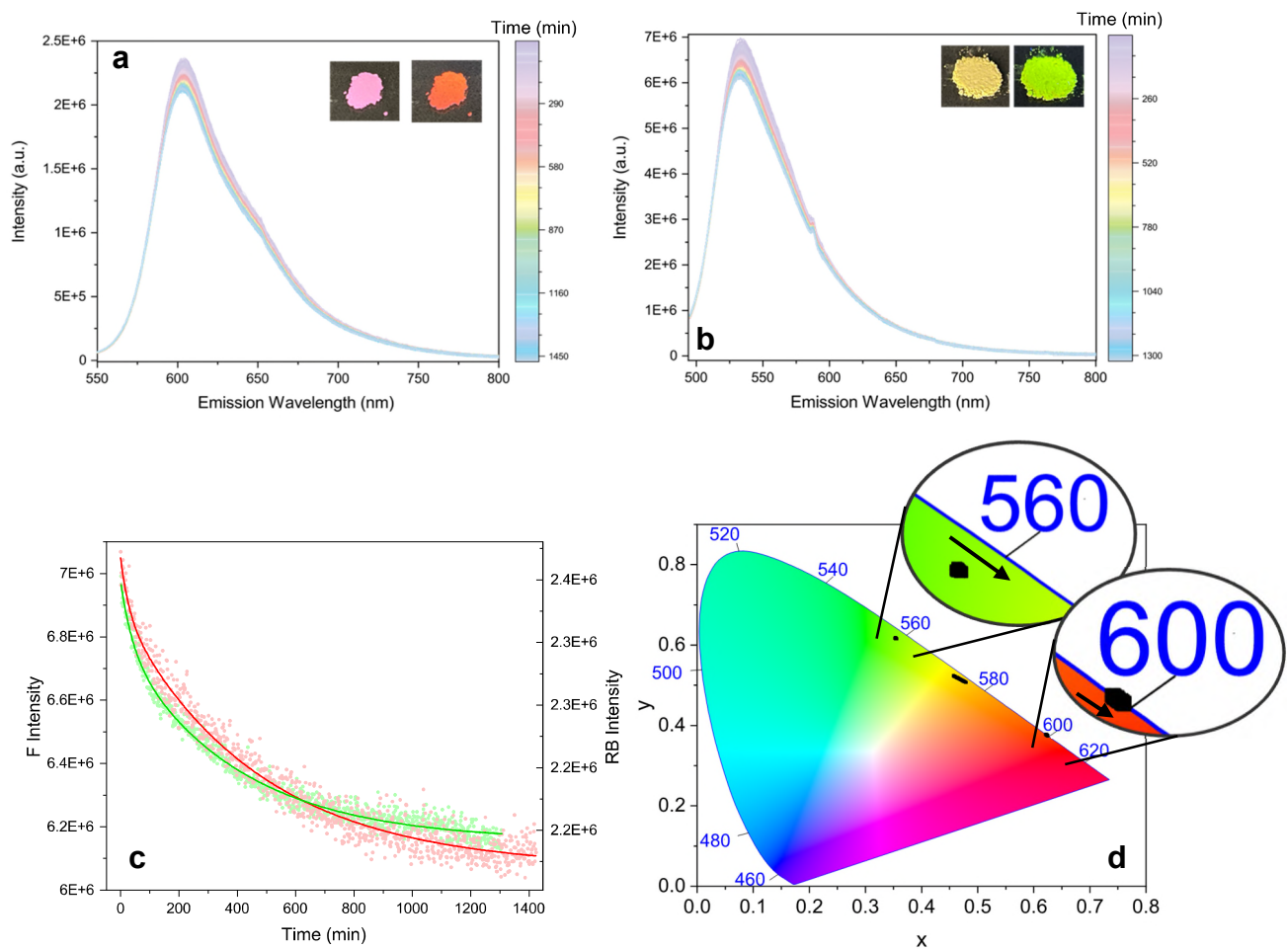

**Figure 25**. Photostability studies of (a) RB@Z7-NS and (b) F@Z7-NS over 24 hour exposure to 150 W xenon bulb emitting at each materials' absorption maximum. Insets show the samples after the 24-hour test subject to ambient light (left) and UV excitation (365 nm). (c) Photostability of various guest@Z7-NS samples as a percentage of maximum absorption peak intensity over 24 hours with continuous exposure to respective sample Abs$_{max}$. (d) emission chromaticity variation of RB@Z7-NS and F@Z7-NS during 24-hour photostability test (left-right).

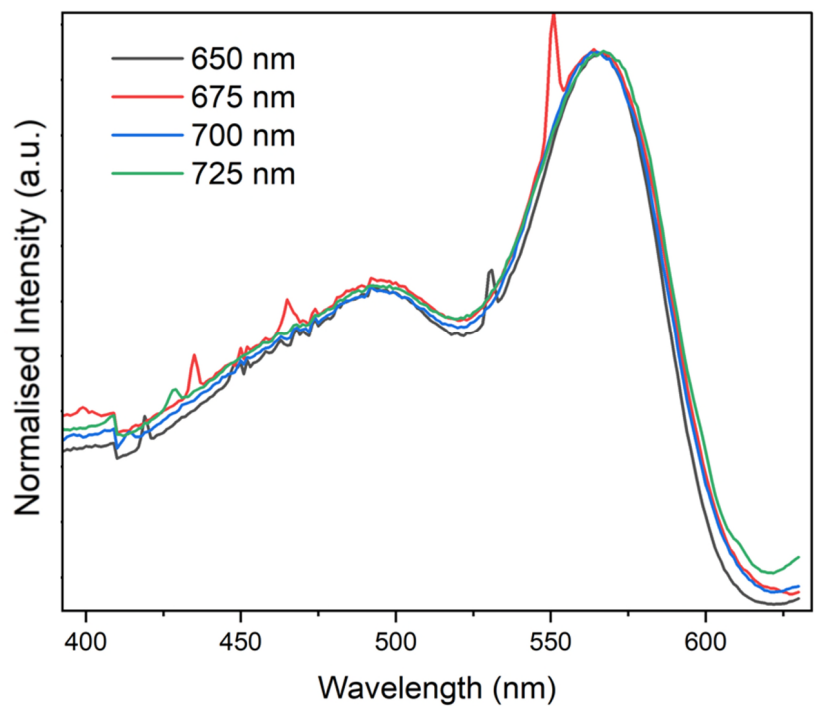

**Figure 26**. Excitation spectra of DG@Z7-NS at various wavelengths (normalised).

| Sample | F Quantity (mmol) | RB Quantity (x100) (mmol) |
|---|---|---|
| 1 | 0.00314 | 0.00165 |
| 3 | 0.0032 | 9.9E-4 |
| 2 | 0.00317 | 0.00132 |
| 4 | 0.00314 | 6.6E-4 |
| 6 | 0.00328 | 0.00247 |
| 5 | 0.00320 | 0.0023 |
| 20 | 0.00327 | 0.00272 |
| 13 | 0.00328 | 0.00247 |
| 17 | 0.00328 | 0.00247 |
| 8 | 0.00329 | 8.25E-4 |
| 19 | 0.00327 | 0.00264 |
| 9 | 0.00328 | 0.00165 |
| 10 | 0.00329 | 0.00132 |
| 14 | 0.00328 | 0.00231 |
| 18 | 0.00327 | 0.00256 |
| 11 | 0.00329 | 9.9E-4 |
| 15 | 0.00328 | 0.00198 |
| 12 | 0.00329 | 8.25E-4 |

**Table 4**. Synthesis concentrations used of F and RB in DG@Z7-NS samples.

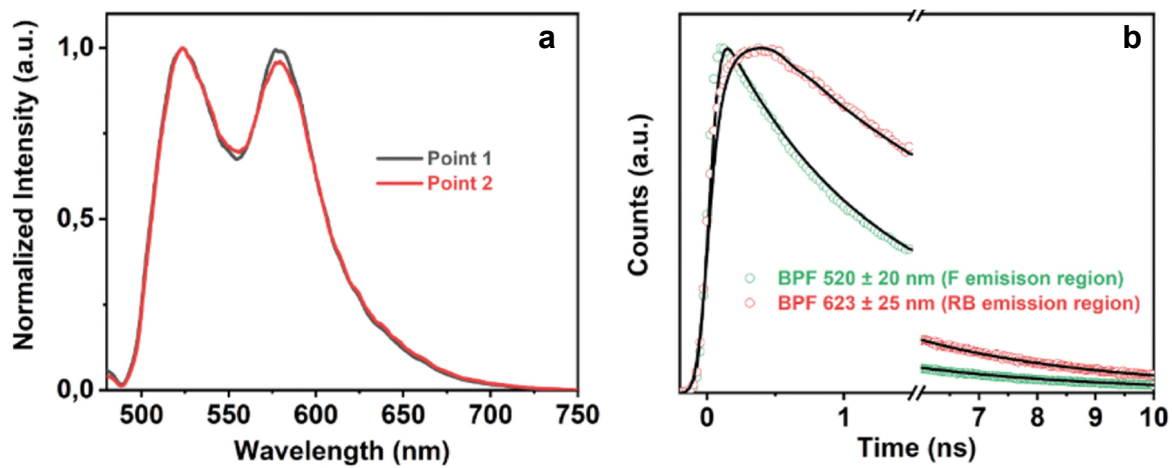

**Figure 27**. DG@Z7-NS emission spectra (a) and lifetime decay curves using different BPF (b) measured with sample under FLIM.

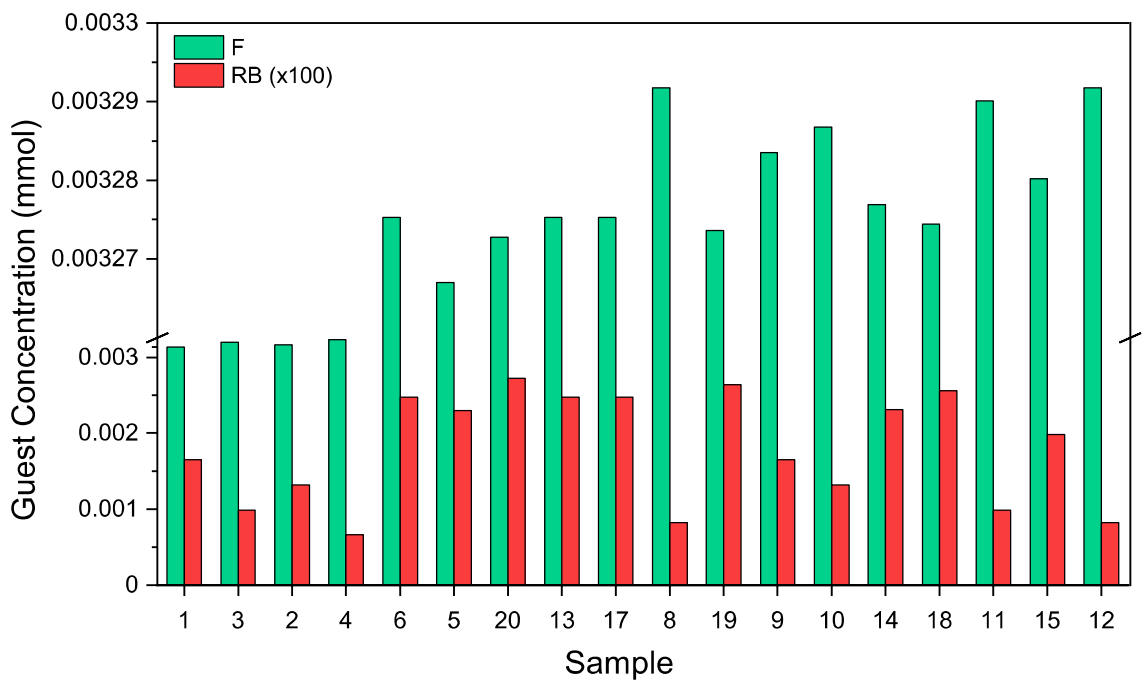

**Figure 28.** Comparison of F and RB content during synthesis of various DG@Z7-NS samples (RB concentration x 100).

| | $\lambda_{obs}$ [nm] | $\tau_1$[ns] | $a_1$ | $c_1$[%] | $\tau_2$[ns] | $a_2$ | $c_2$[%] | $\tau_3$[ns] | $a_3$ | $c_3$[%] | $\chi^2$ |
|---|---|---|---|---|---|---|---|---|---|---|---|
| DG1 | 560 | 0.3823 | 0.0488 | 8.4 | 1.9121 | 0.0603 | 51.87 | 4.3593 | 0.0203 | 39.73 | 1.1217 |
| DG3 | 560 | 0.2538 | 0.0774 | 9.28 | 1.8662 | 0.0606 | 53.45 | 4.4307 | 0.0178 | 37.27 | 1.1777 |
| DG2 | 560 | 0.3753 | 0.0431 | 6.87 | 2.0584 | 0.0605 | 52.91 | 4.6085 | 0.0205 | 40.22 | 1.1372 |
| DG4 | 560 | 0.3689 | 0.0526 | 8.4 | 2.0519 | 0.0619 | 54.96 | 4.7438 | 0.0178 | 36.63 | 1.0234 |
| DG6 | 560 | 0.3518 | 0.0558 | 8.5 | 2.1008 | 0.0616 | 56.09 | 4.7769 | 0.0171 | 35.41 | 1.1728 |
| DG5 | 560 | 0.2289 | 0.0816 | 8.48 | 1.9041 | 0.0621 | 53.71 | 4.4482 | 0.0187 | 37.81 | 1.1298 |
| DG20 | 560 | 0.2336 | 0.081 | 8.35 | 1.9447 | 0.0598 | 51.37 | 4.7699 | 0.0191 | 40.27 | 1.389 |
| DG13 | 560 | 0.412 | 0.0668 | 12.67 | 2.245 | 0.0579 | 59.89 | 5.293 | 0.0113 | 27.44 | 1.2414 |
| DG8 | 560 | 0.3496 | 0.0575 | 8.6 | 2.1482 | 0.0616 | 56.64 | 4.8654 | 0.0167 | 34.75 | 1.1327 |
| DG19 | 560 | 0.1025 | 0.1787 | 7.85 | 2.0151 | 0.0563 | 48.59 | 4.8353 | 0.021 | 43.57 | 1.1836 |
| DG9 | 560 | 0.4014 | 0.054 | 8.59 | 2.3433 | 0.0568 | 52.77 | 5.3684 | 0.0182 | 38.64 | 1.1355 |
| DG10 | 560 | 0.2956 | 0.0689 | 8.89 | 2.0365 | 0.0569 | 50.62 | 4.7746 | 0.0194 | 40.49 | 1.2174 |
| DG14 | 560 | 0.3442 | 0.0439 | 5.85 | 2.2441 | 0.0657 | 57.06 | 4.9995 | 0.0192 | 37.09 | 1.1287 |
| DG18 | 560 | 0.3182 | 0.0616 | 8.6 | 1.9801 | 0.0608 | 52.86 | 4.5705 | 0.0192 | 38.53 | 1.1705 |
| DG11 | 560 | 0.2601 | 0.0635 | 6.21 | 2.0972 | 0.0562 | 44.37 | 5.0728 | 0.0259 | 49.42 | 1.1645 |
| DG12 | 560 | 0.2629 | 0.0768 | 9.09 | 1.9586 | 0.0561 | 49.42 | 4.7151 | 0.0196 | 41.49 | 1.1083 |
| DG15 | 560 | 0.3506 | 0.0409 | 5.35 | 2.3487 | 0.0638 | 55.9 | 5.3481 | 0.0194 | 38.75 | 1.1701 |
| Mix 1:1 (F@Z7-NS to RB@Z7-NS) | 560 | 0.3603 | 0.0786 | 16.23 | 1.8494 | 0.0511 | 54.12 | 4.5427 | 0.0114 | 29.65 | 1.1592 |
| Mix 10:1 (F@Z7NS:RB@Z7-NS | 560 | 0.2049 | 0.1164 | 13.7 | 1.5712 | 0.0531 | 47.96 | 4.1036 | 0.0163 | 38.34 | 1.103 |

**Table 5**. Values of time constants ($\tau_i$), normalised pre-exponential factors ($a_i$), and fractional contributions ($c_i = \tau_i \cdot a_i$) of the emission decay of DG@Z7-NS samples with differing F:RB synthesis concentrations in solid state excitation at 362.5 nm.

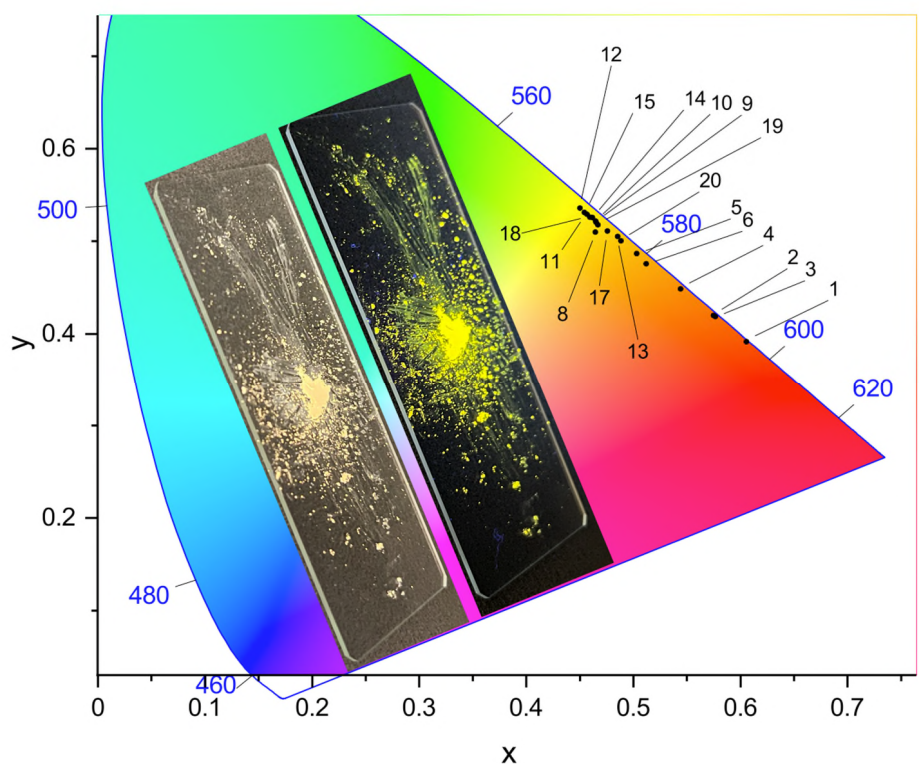

**Figure 29**. Emission chromaticity of various DG@Z7-NS samples. Inset: DG@Z7-NS sample (13) under ambient conditions (left) and UV light (right) emitting warm yellow light.

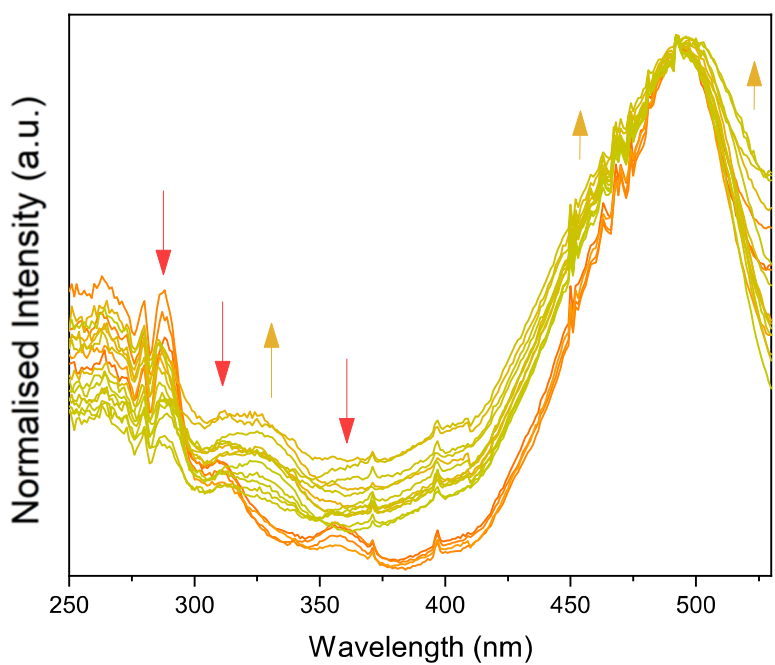

**Figure 30**. Excitation spectra of DG@Z7-NS synthesised with various F:RB concentration. Key spectral variations indicated with arrows.

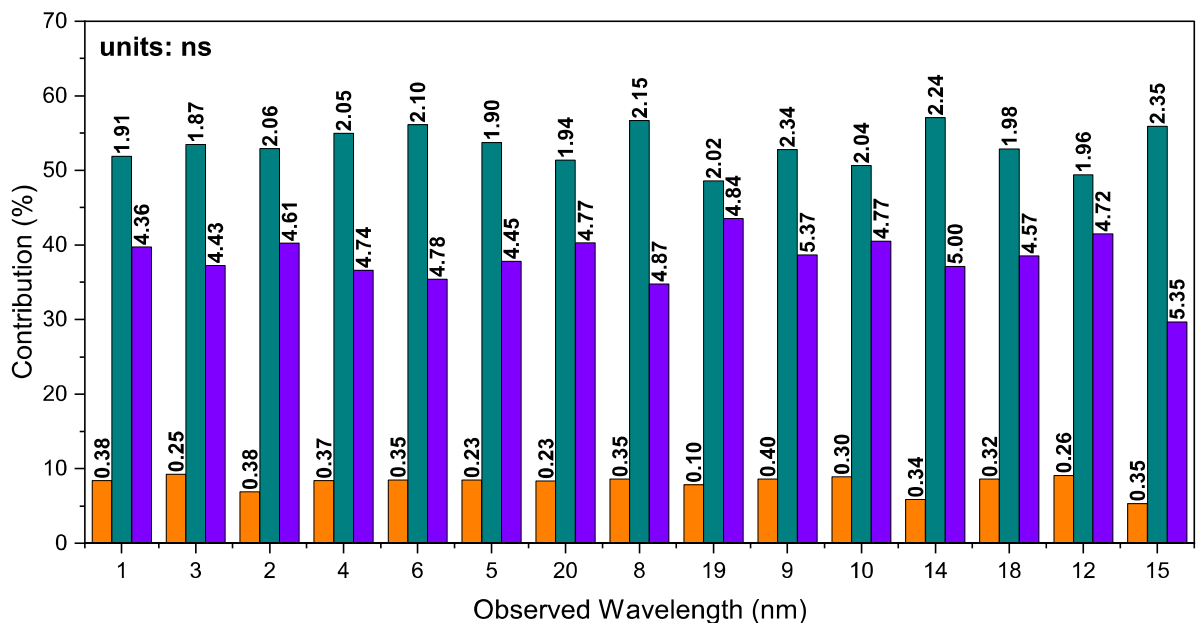

**Figure 31**. Fluorescence Decay Lifetimes of various DG@Z7-NS samples of varying F:RB synthesis concentration (Orange = FRET; green = F and purple = RB).

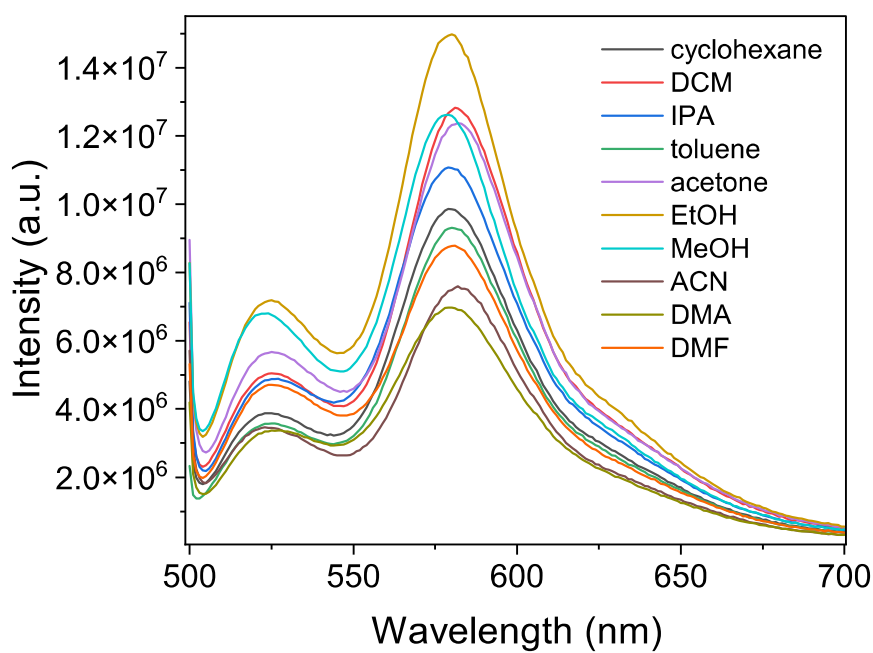

**Figure 32**. Emission spectra of DG@Z7-NS in varying organic solvents.

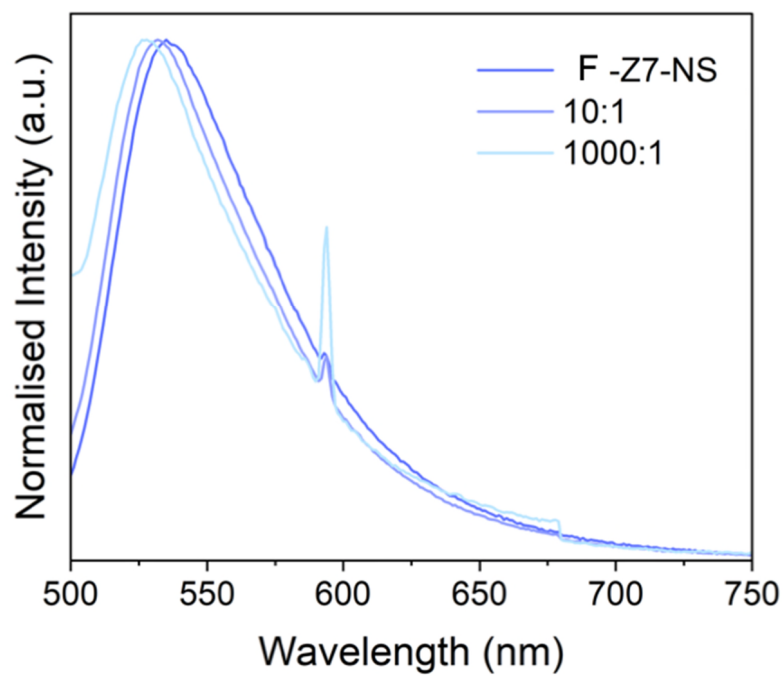

**Figure 33**. Emission spectra of F@Z7-NS when mixed with BaSO₄ in a ratio of 10:1 (BaSO₄:F@Z7-NS) (a) and 1000:1 (b).

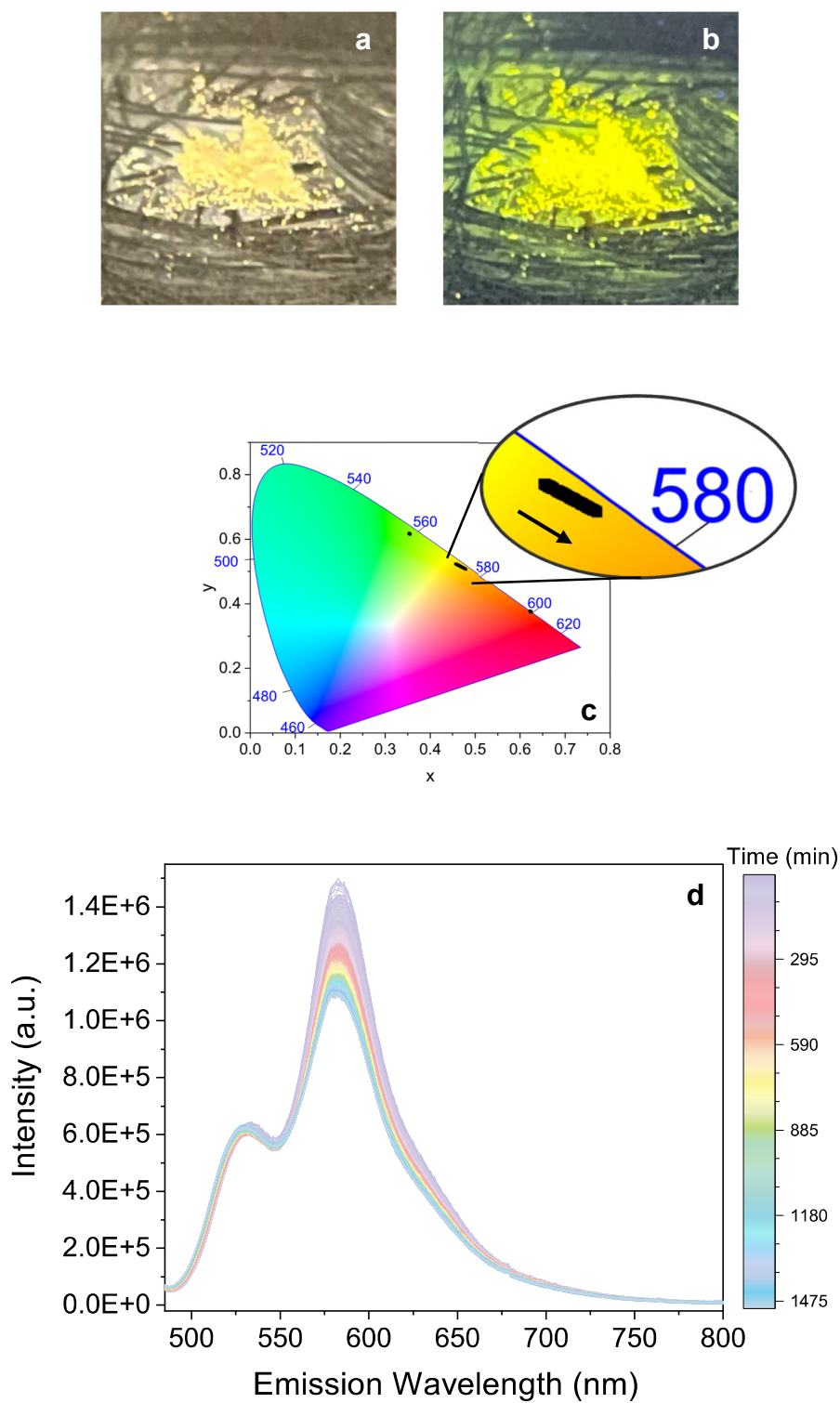

**Figure 34**. Photostability study of DG@Z7-NS over 24-hour exposure to 150 W xenon bulb emitting at the materials' absorption maximum. (a) sample after the 24-hour test subject to ambient light (left) and UV excitation (365 nm) (b). (c) emission chromaticity variation of RB@Z7-NS and F@Z7-NS during 24-hour photostability test (left-right). (d) emission spectra of sample over 24 hours.